\documentclass[10pt,journal,compsoc]{IEEEtran}
\usepackage[utf8]{inputenc}
\usepackage{graphicx}
\usepackage{epstopdf}
\usepackage{float,url}
\usepackage[subrefformat=parens,labelformat=parens]{subfig}
\usepackage{amsmath,epsfig,psfrag,amssymb} 
\usepackage{amsfonts,graphics}
\usepackage{algorithm}
\usepackage{algorithmicx,algpseudocode} 
\usepackage{textcomp}
\usepackage{xcolor}
\usepackage[nocompress]{cite}
\usepackage{multirow}
\usepackage{diagbox} 
\usepackage{bigstrut}
\usepackage{tabularx}
\usepackage{makecell}
\newtheorem{definition}{Definition}
\newtheorem{insight}{Insight}
\newcolumntype{C}[1]{>{\centering\let\newline\\\arraybackslash\hspace{0pt}}m{#1}}

\newcommand{\A}{{\ensuremath{\cal A}}}
\newcommand{\D}{{\ensuremath{\cal D}}}

\newcommand{\aec}{{\ensuremath{\sf AEC}}}

\newcommand{\DI}{{\ensuremath{\sf DI}}}
\newcommand{\AI}{{\ensuremath{\sf AI}}}

\newcommand{\asd}{{\ensuremath{\sf ASD}}}
\newcommand{\awd}{{\ensuremath{\sf AWD}}}
\newcommand{\VT}{{\ensuremath{\sf VT}}}

\newcommand{\fnr}{{\ensuremath{\sf FNR}}}
\newcommand{\fpr}{{\ensuremath{\sf FPR}}}

\newcommand{\tts}{{\ensuremath{\sf TTS}}}

\newcommand{\cc}{{\ensuremath{\sf cc}}}
\newcommand{\oc}{{\ensuremath{\sf OC}}}
\newcommand{\aoc}{{\ensuremath{\sf AOC}}}

\newcommand{\vc}{{\ensuremath{\sf vc}}}
\newcommand{\ic}{{\ensuremath{\sf ic}}}

\newcommand{\APP}{{\ensuremath{\sf APP}}}
\newcommand{\OS}{{\ensuremath{\sf OS}}}
\newcommand{\SW}{{\ensuremath{\sf SW}}}

\newcommand{\VUL}{{\ensuremath{\sf VUL}}}

\newcommand{\app}{{\ensuremath{\sf app}}}
\newcommand{\os}{{\ensuremath{\sf os}}}
\newcommand{\InitialCompromise}{{\ensuremath{\sf IniComp}}}

\newcommand{\ignore}[1]{}

\begin{document}
\title{Quantifying Cybersecurity Effectiveness of Dynamic Network Diversity}

\author{Huashan Chen, Hasan Cam, and Shouhuai Xu
\thanks{H. Chen is with the Department of Computer Science, University of Texas at San Antonio, San Antonio, TX, USA. H. Cam is with Best Buy, Richfield, MN, USA. S. Xu is with the Department of Computer Science, University of Colorado Colorado Springs, Colorado Springs, Colorado, USA; this work was partly done when he was affiliated with University of Texas at San Antonio. Correspondence: {\tt sxu@uccs.edu}}}

\IEEEtitleabstractindextext{%
\begin{abstract}
The deployment of monoculture software stacks can have devastating consequences because a single attack can compromise all of the vulnerable computers in cyberspace. This one-vulnerability-affects-all phenomenon will continue until after software stacks are diversified, which is well recognized by the research community. However, existing studies mainly focused on investigating the effectiveness of software diversity at the building-block level (e.g., whether two independent implementations indeed exhibit independent vulnerabilities); the effectiveness of enforcing network-wide software diversity is little understood, despite its importance in possibly helping justify investment in software diversification. As a first step towards ultimately tackling this problem, we propose a systematic framework for modeling and quantifying the cybersecurity effectiveness of network diversity, including a suite of cybersecurity metrics. We also present an agent-based simulation to empirically demonstrate the usefulness of the framework. We draw a number of insights, including the surprising result that proactive diversity is effective under very special circumstances, but reactive-adaptive diversity is much more effective in most cases.
\end{abstract}

}

\maketitle
\IEEEdisplaynontitleabstractindextext
\IEEEpeerreviewmaketitle

\section{Introduction}
\IEEEPARstart{S}{oftware} monoculture enables the automatic amplification of cyber attack damages because vulnerabilities are replicated network-wide or even cyberspace-wide \cite{Geer2003,Stamp:2004:RM:971617.971650}. As a consequence, a single exploit may allow an attacker to compromise many programs. In order to cope with the problem, researchers have proposed diversifying program implementations \cite{zhang2001heterogeneous,baudry2015multiple,larsen2014sok}, leading to the notion of {\em software diversity}. There are various flavors of  software diversity, such as: $N$-version programming (i.e., a program specification has multiple independent implementations \cite{chen1978n,avizienis1985n}); natural diversity (e.g., multiple browsers incurred by market competition \cite{hiltunen2000survivability});
compiler-based diversification (i.e. random executables generated from a given source code \cite{jackson2011compiler,DBLP:journals/tdsc/HomescuJCBLF17,DBLP:conf/ccs/Franz15,DBLP:conf/ndss/CraneHBLF15,DBLP:journals/ieeesp/LarsenBF14}); and
software runtime environment diversification (e.g., address space layout randomization \cite{bhatkar2003address,forrest1997building,etoh2000gcc,xu2003transparent}, instruction set randomization \cite{kc2003countering,barrantes2003randomized},
system calls randomization \cite{chew2002mitigating}, and replicated execution \cite{salamat2011runtime}). 

It is intuitive that employing software diversity could increase security. For example, the U.S. Navy developed the RHIMES system \cite{wwwonrna56:online} to enhance security of shipboard systems, by introducing diversity to each programmable logic controller. However, real-world software diversity is often employed in an ad hoc fashion, which can be justified by how different OSes (e.g., Windows vs. various kinds of Unix) and browsers (e.g., Safari vs. Firefox vs. Chrome) are employed in practice. One exception is the investigation of employing software diversity to enhance Byzantine Fault-Tolerance (BFT), namely how to employ software diversity in the replica implementations so that they do not contain common vulnerabilities \cite{rodrigues2001base,roeder2010proactive,wood2011zz,platania2014towards,garcia2019lazarus,moniz2008ritas}. This is important because the theoretical fault-tolerance guarantee can be ruined otherwise. Another exception is the investigation of employing software diversity in detecting cyber attacks by leveraging the behavioral discrepancy between  diversified replicas 
\cite{just2002learning,totel2005cots,xu2017platpal}. Despite these studies, some fundamental questions remain open, such as: {\em How should software diversity be employed in practice to amplify, if not maximize, security?}

The preceding question leads to the notion of {\em network diversity}, which deals with the employment of diversified program implementations in network-wide software stacks \cite{Geer2003,ODonnellCCS2004,Stamp:2004:RM:971617.971650,zhang2001heterogeneous,XuComplexNetworkSub2018}. 
A simpler version of the notion may be called {\em static} diversity, where diversified implementations are employed once and for all (i.e., unchanged after initial employment). There are studies on optimizing static diversity via some flavor of {\em graph coloring} algorithm; by treating colors as diversified implementations, the research problem is to minimize defective edges (i.e., adjacent nodes have the same color or run the same implementation) \cite{ODonnellCCS2004,huang2014toward,yang2008improving,XuComplexNetworkSub2018,10.1002/spe.2180,borbor2016diversifying}.
There are also studies on quantifying the effectiveness of {\em static}  diversity \cite{neti2012software,borbor2016diversifying, wang2014modeling,zhang2016network,temizkan2017software}.

Despite these studies, there is no systematic understanding on the network-wide effectiveness of employing software diversity, for multiple reasons. 
First, {\color{black}most studies} use coarse-grained models (i.e., treating each computer as a unit) and do not consider attack-defense interactions. Second, it is not clear how to quantify the network-wide cybersecurity effectiveness of employing network diversity. 
Third, it is not clear how network diversity should be dynamically employed, leading to the notion of {\em dynamic} diversity.
Addressing these problems can deepen our understanding of software diversity, help decision-makers determine whether or not to invest in software diversity, and guide practitioners {\color{black}in intelligently employing} diversified implementations in real-world cyber defense operations.

\smallskip

\noindent{\bf Our contributions}.
We make three contributions.
First, we propose a framework for modeling and quantifying the network-wide cybersecurity of enforcing network diversity. 
The framework considers {\em fine-granularity} by distinguishing applications and operating systems. 
The framework considers the time dimension, which allows us to 
investigate {\em dynamic} diversity; that is, the employment of diversified implementations in the network evolves over time.

Second, we propose a suite of metrics to quantify the network-wide effectiveness of employing diversity, including: (i) {\em time-to-succeed}, which measures how long it takes an attacker to break a defender's goal (if possible); (ii) {\em attacker slow-down}, which measures the extent an attacker is slowed down by network diversity; (iii) {\em attack worst damage}, which measures the damage an attacker can cause in the worst case; (iv) {\em attack extra cost}, which measures the extra investment the defense imposes on an attacker in order to break the defender's goal; (v) {\em vulnerability tolerance}, which measures the upper bound of vulnerabilities that can be tolerated when achieving the defender's goal; (vi) {\em average operational cost}, which measures the average fraction of programs that re-deploy dynamic diversity. These metrics may be of independent value.

Third, we demonstrate the usefulness of the framework and metrics by presenting a multi-agent simulation study with multiple network diversity strategies: monoculture (the baseline with no diversity), static diversity (for comparison purposes), proactive (periodically re-diversifying network-wide software stacks), reactive-adaptive (re-diversifying the network stacks in response to detected attacks), and hybrid (of the last two). According to the simulation study, we draw a number of insights, including: (i) In terms of attacker slow-down, reactive-adaptive diversity is the most effective strategy and the initial diversity configuration matters. (ii) In order to reduce the attack worst damage, different diversity strategies should be used in different parameter regimes. (iii) Reactive-adaptive diversity leads to a higher vulnerability-tolerance than proactive diversity does. (iv) Proactive diversity improves security only when dynamic diversity is widely re-employed at a high frequency, which however incurs a high operational cost.
(v) The more the diversified implementations, the higher the attacker slow-down, the higher the attack extra cost, and the higher the vulnerability tolerance. This is especially true for reactive-adaptive diversity. 
Note that these findings may not be universally true because they are derived from the parameter settings used in the simulation study.

\smallskip

\noindent{\bf Paper outline}. 
Section \ref{sec:framework} presents the framework. Section \ref{sec:case-study} describes the simulation study. Section \ref{sec:related-work} reviews related prior work. Section \ref{sec:limitations} discusses the limitations of the present study. Section \ref{sec:conclusion} concludes the paper.
Table \ref{table:notations} summarizes the notations used throughout the paper.

\begin{table}[!htbp]
\centering
\begin{tabular}{|l|p{.4\textwidth}|}
\hline
$\A$, $\D$ & attacker $\A$ and defender $\D$  \\ \hline
$n,i,j$ & $n$ is the number of computers in a network, $\app_{i,j}$ is the $j$-th application running in computer $i \in [1,n]$\\ \hline
$\ell,z$ & $\ell$ is the number of phases of $\A$'s strategy, $z\in [1,\ell]$\\ \hline
$\hbar,k$ & {$\hbar$ is the number of different programs, $k\in [1,\hbar]$} \\ \hline
$X_k$ & $X_k$ is the number of diversified implementations of program $k$ \\ \hline
$Q_k$ & {software quality of diversified implementations of the $k$-th program (functionality), $Q_k\in [0,1]$ interpreted as probability of containing vulnerability }\\ \hline
$G$	& $G=(V,E)$ is the communication graph of a network (rather than the network's physical topology), where $v\in V$ represents a program (application or operating system)\\
\hline
$C_t$ & $C_t:V\to \SW$ is the network diversity configuration at time $t$, where $\SW$ is the set of diversified implementations of all programs\\
\hline
$\phi_t$ & $\phi_t: C_t(V)\to 2^{\VUL}$ is the mapping from diversified programs to the vulnerabilities present in them at time $t$. $\Phi_t = (\phi_t(C_t(v)))_{v\in V}$ is the vulnerabilities in the network.
\\
\hline
$s_{v,t}$    & $s_{i,t}\in \{0,1,2\}$ is the cybersecurity state of node (i.e., program running at) $v\in V$ at time $t$ (0: {\em vulnerable}; 1: {\em compromised}; 2: {\em invulnerable}). Vector $S_t=(s_{v,t})_{v\in V}$ \\ \hline
$s_{i,t}$    & $s_{i,t}\in \{0,1,2\}$ is the cybersecurity state of computer $i$ at time $t$, similarly defined as $s_{v,t}$. Vector $S'_t=(s_{i,t})_{i\in [1,n]}$ \\ \hline
$\Sigma_t$ & $\Sigma_t = (G, C_t, \Phi_t, S_t)$ is 
the cybersecurity situation of a network at time $t$ \\
\hline
$\Sigma_{\A,t}$ & $\Sigma_{\A,t}=(G_{\A,t}, C_{\A,t}, \Phi_{\A,t}, S_{\A,t})$ is $\A$'s perception of the target network
$\Sigma_t$ at time $t$ \\
\hline
$\Omega_{\A,t}$ & $\Omega_{\A,t}=(\omega_{v,\A,t})_{v\in V_{\A,t}}$ is $\A$'s goal at time $t$ at the {\em program} level, where $\omega_{v,\A,t}\in \{\bot,1\}$; $\Omega'_{\A,t}=(\omega_{i,\A,t})_{i\in [1,n]}$ is $\A$'s goal at time $t$ at the {\em computer} level, where $\omega'_{i,\A,t}\in \{\bot,1\}$ \\
\hline
$\Gamma_\A$ & $\Gamma_\A = \{\gamma_{\A,1},\ldots,\gamma_{\A,\ell}\}$ is attacker's strategy of $\ell$ phases \\
\hline
$\Delta_\A$ & $\Delta_\A=(\Delta_{\A,z})_{z\in [1,\ell]}$ is an attacker's capability where $\Delta_{\A,z} = \{\psi_{\A,z,1}, \ldots,\psi_{\A,z,m_z}\}$ is the exploits that are applicable and available to the attacker at phase $z$ \\
\hline
${\cal F}_{\A,t}$ & ${\cal F}_{\A,t}$ is the attacker's decision-making algorithm  to make an attack plan $\Lambda_\A=(\lambda_{\A,1},\ldots,\lambda_{\A,\ell})$ at time $t$, where $\lambda_{\A,z}\in \Delta_{\A,z}$ for $1\leq z\leq \ell$\\
\hline
$\Sigma_{{\D},t}$ & $\Sigma_{{\D},t}=(G_{\D},C_{{\D},t}, \Phi_{{\D},t}, S_{{\D},t})$ is $\D$'s
perception of $\Sigma_t$ at time $t$ \\
\hline
$\Omega_{\D,t}$ & $\Omega_{\D,t}=(\omega_{v,\D,t})_{v\in V}$ is $\D$'s goal at time $t$ at the {\em program} level, where $\omega_{v,\D,t} \in [0,1]$; $\Omega'_{\D,t}=(\omega_{i,\D,t})_{i\in [1,n]}$ is $\D$'s goal at time $t$ at the {\em computer} level, where $\omega_{i,\D,t} \in [0,1]$\\
\hline
$\Gamma_{\D}$ & $\Gamma_\D = \{\gamma_{\D,1},\ldots,\gamma_{\D,\jmath}\}$ is defender's strategy \\
\hline
$\eta_{\D,1}$ & $\eta_{\D,1}$ represents the proportion of nodes in $V$ employing  diversified implementations \\
\hline
$\eta_{\D,2}$ & $\eta_{\D,2}$ represents the {\em frequency} at which the diversified implementations will be dynamically re-employed \\
\hline
$\eta_{\D,3}$ & $\eta_{\D,3}$ represents the {\em condition} under which diversified implementations are re-employed \\
\hline
$\Delta_\D$ & $\Delta_\D=(\Delta_{\D,k})_{k\in [1,\hbar]}$ is defender's capability, where $\Delta_{\D,k} = \{\delta_{\D,k,1}, \ldots,\delta_{\D,k,X_k}\}$ is a set of $X_k$ diversified implementations of program $k$;
\\
\hline
${\cal F}_{\D,t}$ & the decision-making algorithm to make a defense plan $\Lambda_{\D,t}=(\delta_{\D,t}(v))_{v\in V}$ at time $t$ \\
\hline
$\tau$  & the compromise probability of computer $i\in[1,n]$ at time $t\in[0,T]$ that can be tolerated by the defender  \\
\hline
$\tts$ & $\tts(\A,\D)=\min\{t:\cc(t)> \tau\}$ measures how long it takes for attacker $\A$ to break defender $\D$'s mission goal $\tau$\\
\hline
$\asd$ & $\asd_{\D,q}=\tts(\A,\D_q)-\tts(\A,\D_1)$ measures the extent at which $\A$ is slowed down by {defense strategy $\gamma_{\D,q}$} \\
\hline
$\awd$ & $\awd_{G,\hbar,X,Q}(\A,\D, T) = \max\{\cc(t):t\leq T\}$ measures attack worst damage during mission lifetime $t\in[0,T]$\\
\hline
$\aec$ & $\aec_{\D,q}=\AI(\gamma_{\D,q})-\AI(\gamma_{\D,1})$ measures the number of extra exploits $\A$ needs to obtain to make $\awd_{G,\hbar,X,Q}(\A,\D_q, T) > \tau$ against {defense strategy $\gamma_{\D,q}$}\\
\hline
$\VT$ & $\VT_{\D,q} = \max\{Q:\awd_{G,\hbar,X,Q}(\A,\D_q, T) {\leq} \tau\}$ captures the upper bound of tolerable vulnerabilities such that $\D$ can still achieve its mission goal $\tau$ in lifetime $[0,T]$\\
\hline
$\aoc$ & $\aoc_{\D,q}(T)=\sum_{t=1}^{T} \oc_{\D,q}(t)/T$ is the average operational cost to achieve the defender's goal; $\aoc_{\D,q}^{\sf max} = $ $ \max\{\aoc_{\D,q}(T):\awd_{G,\hbar,X,Q}(\A,\D_q, T)\leq \tau\}$ and $\aoc_{\D,q}^{\sf min} = \min\{\aoc_{\D,q}(T):\awd_{G,\hbar,X,Q}(\A,\D_q, T)\leq \tau\}$ are maximum and minimum $\aoc_{\D,q}(T)$, respectively.\\
\hline
\end{tabular}
\vspace{-2mm}
\caption{Summary of main notations.\label{table:notations}}
\end{table}

\begin{figure*}[!htbp]
\centering
\includegraphics[width=.8\textwidth, height = .25\textheight]{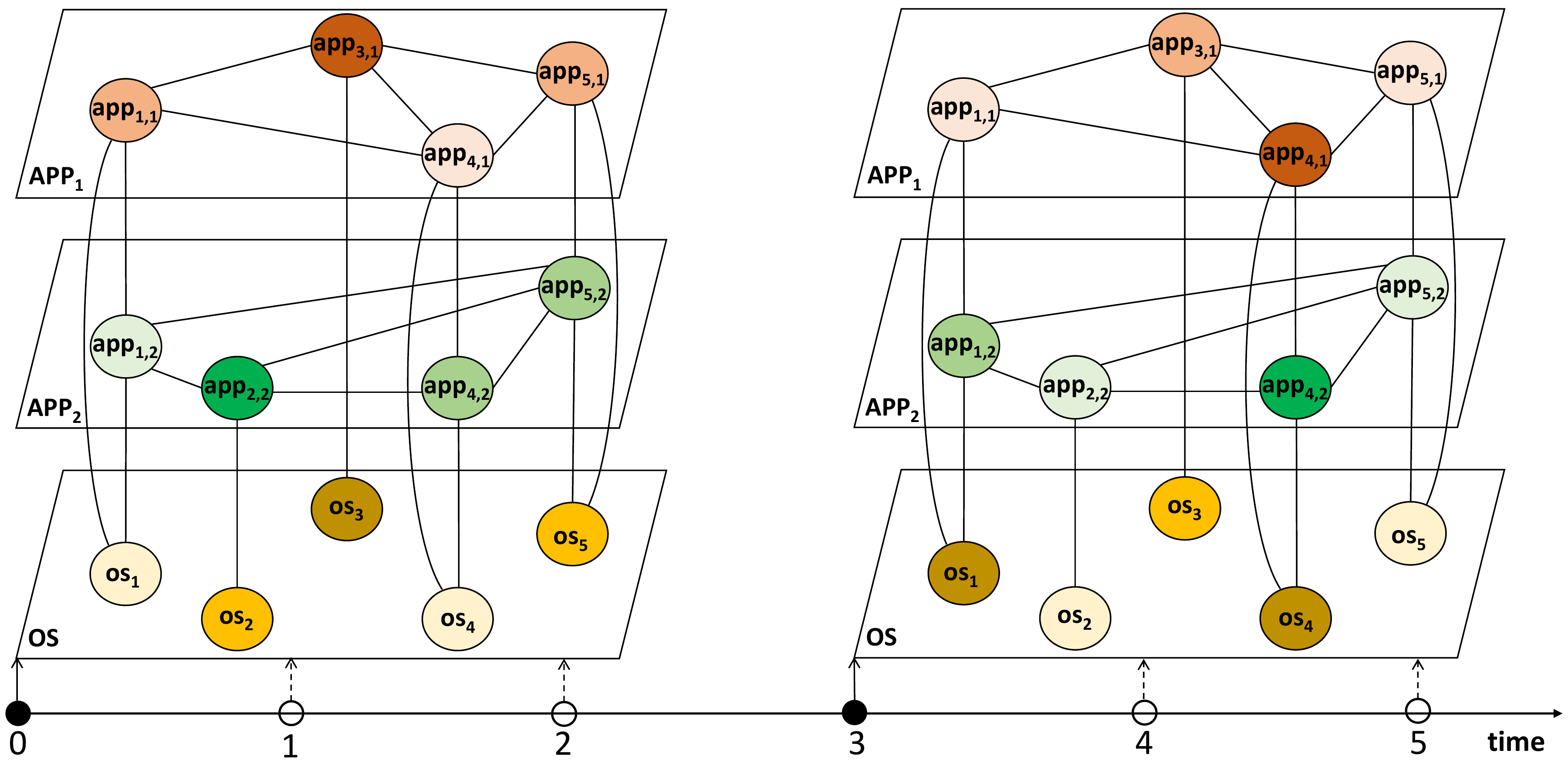}
\caption{{\color{black}
Illustration of dynamic diversity in a network of 5 computers ($1\leq i \leq 5$). The time horizon shown is $t=0,1,\ldots,5$.}
\label{fig:dynamic-diversity-illustration}}
\end{figure*}

\section{The Framework} 
\label{sec:framework}

Consider a network, where each node represents a computer. Each computer has a software stack, running an operating system (OS) in the kernel space and some application(s) in the user space.
Each {\em program}, OS and application alike, {\color{black}may have diversified implementations (e.g., Safari vs. Firefox vs. Chrome for browser)}. These diversified implementations may be obtained by using some of the diversification methods mentioned above (e.g., $N$-version programming or natural diversity via market competition). Each program may or may not be vulnerable.

\subsection{Intuition of Dynamic Network Diversity}

Given diversified implementations of programs, the aforementioned notion of {\em static} network diversity is to employ diversified programs in the computers' software stacks, where the employment (or configuration) will not change during the time horizon of interest. 
We initiate the aforementioned {\em dynamic} network diversity, which aims to dynamically employ diversified programs at the computers' software stacks, where the employment (or configuration) does change during the time horizon of interest. 

{\color{black}
Figure \ref{fig:dynamic-diversity-illustration} illustrates the idea via a network of 5 computers, denoted by $1\leq i \leq 5$. The time horizon is $t=0,\ldots,5$. Each computer runs an OS, and we use $\os_i$ to represent the OS running in computer $i$. There are two applications, denoted by {$\APP_j$} for $1\leq j \leq 2$ (e.g., $\APP_1$ is browser and $\APP_2$ is email). Computers 1, 4 and 5 run both applications; computer 2 runs $\APP_2$; computer 3 runs $\APP_1$. We use $\app_{i,j}$ to represent the application $\APP_j$ running in computer $i$. {\bf (i)} There are three implementations of OS (e.g., Windows vs. Linux vs. macOS), which are respectively indicated by three colors. For example, at time $t=0,1,2$, {\color{black}computers 1 and 4 run the same OS, causing $\os_1$ and $\os_4$ to have the same color; computers 2 and 5 run another OS, causing $\os_2$ and $\os_5$ to have another color; computer 3 runs yet another OS, causing $\os_3$ to have a different color.}
{\bf (ii)} There are three implementations of $\APP_1$ (e.g., Safari vs. Firefox vs. Chrome), which are respectively indicated by three colors. For example, {\color{black}at time $t=0,1,2$, computers 1 and 5 run the same implementation of $\APP_1$, causing $\app_{1,1}$ and $\app_{5,1}$ to have the same color; computers 3 and 4 run two other implementations of $\APP_1$, causing $\app_{3,1}$ and $\app_{4,1}$ to have different colors;} computer 2 does not run $\APP_1$.
{\bf (iii)} There are three implementations of $\APP_2$ (e.g., Outlook vs. Thunderbird vs. eM Client), which are respectively indicated by three colors. 
For example, {\color{black}at time $t=0,1,2$, computers 4 and 5 run the same implementation of $\APP_2$, causing $\app_{4,2}$ and $\app_{5,2}$ to have the same color; computers 1 and 2 run two other implementations of $\APP_2$, causing $\app_{1,2}$ and $\app_{2,2}$ to have different colors;}
computer 3 does not run $\APP_2$.

Figure \ref{fig:dynamic-diversity-illustration} illustrates dynamic diversity as follows.  At time $t=0$, the network-wide diversity is configured to run a certain combination of specific implementations of OS and applications as indicated by colors. The solid arrow at $t=0$ indicates that a new (i.e., initial) diversity configuration is employed. This configuration remains unchanged for $t=1,2$, as indicated by dashed arrows at $t=1,2$. 
At time $t=3$, the network-wide diversity is re-configured to run another combination of the diversified  implementations of OS and applications. The solid arrow at $t=3$ indicates this new employment. The configuration remains unchanged for $t=4,5$, as indicated by dashed arrows at $t=4,5$.
}

\subsection{Framework Overview}

\begin{figure*}[!htbp]
\centering
\includegraphics[width=1.0\textwidth]{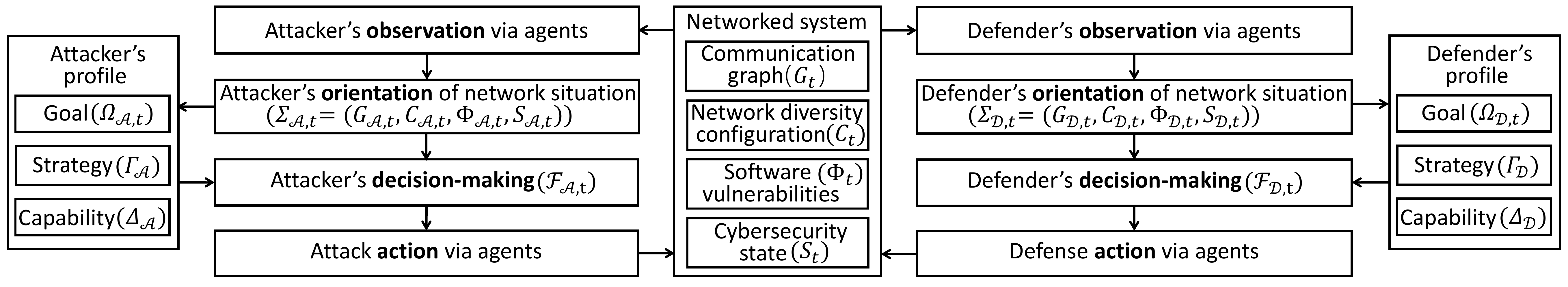}
\caption{The COODAL framework for characterizing the effectiveness of dynamic network diversity.
\label{fig:diversity-model}}
\vspace{-0.3cm}
\end{figure*}

Figure \ref{fig:diversity-model} highlights the Cyber Observation-Orientation-Decision-Action Loop (COODAL) framework for describing attack-defense interactions in a network. COODAL adapts the military operation concept of OODA Loop \cite{BoydOODALOOP1995} to cyberspace. At a high level, the network is abstracted as a communication graph. Both attacker and defender have their own Observation (collecting data), Orientation (analyzing the collected data while leveraging relevant intelligence or information), Decision (determining what to do),  and Action (executing the decision). We consider a discrete-time model with a finite time horizon $t=0,1,\ldots,T$, where $T$ is the defender's mission lifetime (i.e., certain security requirements must be satisfied in order to achieve mission assurance in time interval $[0,T]$). For simplicity, we assume an attack or defense action takes effect instantly (if effective). 
When desired, this assumption can be eliminated by explicitly modeling the delay for an action to take effect (e.g., the framework can be extended to accommodate that an attack or defense action occurs at time $t$ and takes effect at $t+1$).

We stress that the framework does not make restrictive assumptions. For example, we do not make any restriction on how vulnerabilities may be distributed among the diversified implementations {\color{black}by equally accommodating the scenarios where diversified implementations may or may not have common vulnerabilities.} When the attacker has an exploit against a vulnerability that is common to multiple programs, the attacker can compromise all of these vulnerable programs. {\color{black}A successful attack against an OS implies a successful attack against the applications running on top of the OS; a successful attack against an application paves a way for the attacker to attack the OS beneath the application (e.g., privilege escalation).}

\subsection{Modeling Network}
\noindent{\bf Modeling computers' software stacks}. 
A network (e.g., enterprise or mission network) consists of $n$ interconnected computers or devices. Each computer runs a {\em software stack}, which has two layers: application and operating system (OS). We treat any software running in the {\em user space} as application program. By contrast, an OS runs in the {\em kernel space}.
{\color{black}Consistent with the notations used above, we use $\APP$ to denote the universe of application programs. As mentioned above, computer $i$ runs one or multiple applications, denoted by $\app_{i,j}$, where $1\leq i \leq n$, $j$ is the index of the application, and $\app_{i,j} \in \APP$. We use $\OS$ to denote the universe of OSes, and use $\os_i$ to denote the specific OS running in computer $i$, where $\os_i \in \OS$.}
Let $\hbar$ denote the number of different programs running in the network.

\smallskip

\noindent{\bf Modeling communications}. 
We explicitly model the communications between applications because they can be leveraged by the attacker to wage attacks. There are two types of communications: {\em intra-computer} and {\em inter-computer} \cite{XuHotSoS2018Firewall}.
Intra-computer communications are conducted by the programs running in a computer and can be represented as {\em edges} in the terminology of Graph Theory.  
For example, Figure \ref{fig:dynamic-diversity-illustration} shows that the two applications running in computer 1 (i.e., $\app_{1,1}$ and $\app_{1,2}$) are designed to communicate with each other, and that both $\app_{1,1}$ and $\app_{1,2}$ can communicate with $\os_1$ (e.g., for making system calls).
Inter-computer communications are conducted by applications running in different computers and can be represented as {\em edges}. Note that OSes may not communicate with each other. We distinguish intra-computer and inter-computer communications for two reasons: (i) they are often leveraged by the attacker for different purposes --- the former for privilege escalation and the latter for lateral movement between computers; and (ii) they are defended by different security mechanisms --- the former is defended by host-based mechanisms (e.g., intrusion prevention) and the latter is defended by network-based mechanisms (e.g., firewall).

Formalizing the preceding discussion, we naturally obtain the notion of {\em communication graph} $G=(V,E)$, where each vertex or node $v\in V$ represents (and runs) a program and each edge $(u,v)\in E$ represents that a pair of nodes are permitted to communicate with each other. Note that a computer is represented by a set of nodes in $G$ because it runs a set of programs. Since {\em programs} run at {\em nodes} in $G$, we use {\em programs} and {\em nodes} interchangeably to make succinct statements.
We use the term {\em edges} in the standard way to indicate {\em undirected} graphs; a communication graph can be {\em directed} in principle, which is however rare in practice. In the example illustrated in Figure \ref{fig:dynamic-diversity-illustration}, where have $|V|=13$ (i.e., 4 nodes running $\APP_1$, 4 nodes running $\APP_2$, and 5 nodes running OS), and the edge set $E$ is as illustrated. 

We stress that communication graph $G=(V,E)$ is different from a networking-induced graph, for two reasons. First, a vertex or node in a communication graph represents a program running in a computer and each edge represents the communication between two programs. In contrast, a node in a networking-induced graph often represents a computer, which can run multiple programs.
Second, a communication graph can encode access control policies, which may regulate which programs are (not) allowed to communicate with which other programs running in the same computer or different computers. This means that a communication graph may not be a complete graph because some programs may only be allowed to communicate with some of the others. Whereas, a networking-induced graph cannot encode access control policies and would be a complete graph because any computer can communicate with any other computer as long as they are routable.

In this paper we assume a communication graph $G=(V,E)$ is {\em time-independent}, meaning that the applications running in a computer are fixed. This means that the applications running in a computer do not change, but their specific implementations may change over time. This is plausible because we focus on quantifying the effectiveness of dynamic network diversity. 
Nevertheless, $G=(V,E)$ can be extended to time-dependent $G_t=(V_t,E_t)$ when desired.

\smallskip
\noindent{\bf Modeling network diversity configuration}. 
Let $\SW$ denote the set of diversified implementations of the $\hbar$ programs running in the network, including applications and OSes. The {\em dynamic} network diversity configuration at time $t$ refers to the mapping from the set of programs to their specific implementations. This mapping can be described by a function $C_t:V\to \SW$ such that $C_t(v)$ is a specific implementation of the program running at node $v\in V$ at time $t$. For example, Figure \ref{fig:dynamic-diversity-illustration} shows that there are 3 different programs (i.e., $\APP_1$, $\APP_2$, OS); each program has 3 diversified implementations (indicated by 3 different colors) and thus $|\SW|=9$, where computer $1$ runs a specific implementation of each of the three programs at $t\in [0,2]$.
Note that the preceding representation is general enough to accommodate the {\em static} network diversity, which corresponds to $C_0=C_1=\ldots=C_T$, and the {\em monolithic} software stack, which corresponds to each program having exactly one implementation (i.e., $|\SW|$ equals the number of different programs and $C_0=C_1=\ldots=C_T$).

\smallskip

\noindent{\bf Modeling software vulnerabilities}. 
The software running in a computer, application and OS alike, may contain vulnerabilities. We use $\VUL$ to denote the universe of software vulnerabilities that may be present in the software stacks running in the computers of the network in question. Since vulnerabilities are associated with the nodes in $V$, {we use function
$\phi_t: C_t(V)\to 2^{\VUL}$ 
to describe the set of vulnerabilities that are present in the implementations of application and OS programs running in the computers, where $\phi_t(C_t(v))=\emptyset$ means the implementation running at node $v$ is not vulnerable.} 
We use vector $\Phi_t = (\phi_t(C_t(v)))_{v\in V}$
to denote the {\em ground-truth} vulnerabilities that are present in the software running in computers at time $t$. 
Note that this ground-truth vulnerability set may or may not be known to the attacker or defender, and that defining $\phi_t: C_t(V)\to 2^{\VUL}$ is equivalent to defining it as a function from the software set $\SW$ to $2^{\VUL}$. We choose the former because it simplifies subsequent discussion (in regards to dynamic network diversity).
Let $Q_k\in [0,1]$, $1\leq k \leq \hbar$, denote the quality of program, which is measured by the ratio of the number of vulnerable implementations to the total number of diversified implementations of program $k$. The parameter $Q_k$, which is interpreted as the probability that program $k$ is vulnerable, would depend on the technologies that are employed to reduce vulnerabilities in the course of software development (e.g., vulnerability detection \cite{XuVulPeckerACSAC2016,VulDeePecker,DBLP:journals/corr/abs-2001-02334,XuSySeVR2018,XuVulDeeLocator}). The smaller the $Q_k$, the higher the diversity quality of program $k$.

\smallskip

\noindent{\bf Modeling network-wide cybersecurity state and situation}. 
Cybersecurity state and situation can be defined at the {\em program} of $v\in V$ and at the computer level. The former is suitable for modeling and simulation purposes, and the latter is more suitable for cyber defense operation and management purposes. So, we consider both. 

At the program level, we consider a communication graph $G=(V,E)$, which abstracts a network as described above. We use $s_{v,t}\in \{0,1,2\}$ to denote the state of node $v \in V$ at time $t$: $s_{v,t}=0$ means $v$ (i.e., the program running at $v$) is {\em vulnerable} but not compromised, namely that $v$ contains a vulnerability but the vulnerability is not exploited by the attacker and the underlying OS is not compromised; $s_{v,t}=1$ means $v$ is {\em compromised} {either because its vulnerability is exploited or because the underlying OS is compromised}; and $s_{v,t}=2$ means $v$ is {\em invulnerable} (i.e., containing no vulnerabilities) and the underlying OS is not compromised. The {\em program-level} network-wide {\em cybersecurity state} at time $t$ is represented by vector $S_t=(s_{v,t})_{v\in V}$. 

{At the computer level, we say computer $i$ is {\em vulnerable} (or $s_{i,t}=0$) if any program running in the computer is vulnerable, {\em compromised} (or $s_{i,t}=1$) if any program running in the computer is compromised, and {\em invulnerable} (or $s_{i,t}$=2) if all programs running in the computer are invulnerable. Similarly, we can define {\em computer-level} network-wide cybersecurity state at time $t$ as $S'_t=(s_{i,t})_{i\in [1,n]}$, where $S'_t$ can be derived from $G$ and $S_t$.
We further define the vector  ($(\vc(t),\cc(t),\ic(t))$) to succinctly describe the computer-level network-wide cybersecurity effect at time $t$, where $\vc(t)=|\{i:s_{i,t}=0\}|/n$ is the fraction of {\em vulnerable} computers, $\cc(t)=|\{i: s_{i,t}=1\}|/n$ is the fraction of {\em compromised} computers, and $\ic(t)=|\{i:s_{i,t}=2\}|/n$ is the fraction of {\em invulnerable} computers. Note that $\cc(t)+\vc(t)+\ic(t)=1$.} 

The network-wide {\em cybersecurity situation} can be described by $\Sigma_t = (G, C_t, \Phi_t, S_t)$, or a tuple of the communication graph, the network diversity configuration,
the set of vulnerabilities associated with each implementation,
and the network-wide cybersecurity state. {Note that we do not mention $S'_t$ because it can be derived from $S_t$.}

\subsection{Modeling Attacker}

We model an attacker $\A$ (i.e., threat model) with five attributes: {\em knowledge} (what $\A$ knows about a network), {\em goal} (what $\A$ attempts to achieve), {\em strategy} (what strategy $\A$ uses), {\em capability} (what exploits $\A$ possesses), and {\em decision-making} (i.e., what decision-making algorithms $\A$ uses). Intuitively, $\A$ with a certain knowledge attempts to achieve a goal by leveraging some capabilities to compromise some nodes or computers according to some strategies.

\smallskip

\noindent{\bf Attacker's knowledge}.
{We define attacker $\A$'s knowledge as vector $\Sigma_\A=(\Sigma_{\A,t})_{t\in [0,T]}$ such that $\Sigma_{\A,t}=(G_{\A,t}, C_{\A,t}, \Phi_{\A,t}, S_{\A,t})$ is $\A$'s perception of the target network
$\Sigma_t= (G, C_t, \Phi_t, S_t)$ at time $t$,} where $G_{\A,t}=(V_{\A,t}, E_{\A,t}) \subseteq G$ is the attacker's perception of $G$, $C_{\A,t}$ is the attacker's perception of $C_t$, $\Phi_{\A,t}$ is the attacker's perception of $\Phi_{t}$, and $S_{\A,t}$ is the attacker's perception of $S_t$. {\color{black}Note that $S_{\A,t}=S_t$ because the attacker knows which programs are compromised by the attacker itself.} Note also that the notion of {\em initial compromise} can be modeled as part of the attacker's knowledge at time $t=0$, because it describes which nodes $v\in V$ are  compromised at $t=0$. We use $\InitialCompromise = \{v\in V: s_{v,0}=1\}$ to denote the set of  programs that are compromised at $t=0$.

\smallskip

\noindent{\bf Attacker's goal}. \label{definition:attacker-goal}
We define attacker $\A$'s goal as vector $\Omega_\A=(\Omega_{\A,t})_{t\in [0,T]}$ where vector $\Omega_{\A,t}=(\omega_{v,\A,t})_{v\in V_{\A,t}}$ is $\A$'s goal at time $t$, where $\omega_{v,\A,t}\in \{\bot,1\}$, 
$\omega_{v,\A,t}=\bot$ means that $\A$ does not care about the state of $v\in V_{\A,t}$ at time $t$, $\omega_{v,\A,t}=1$ means $\A$ attempts to make $v\in V$ compromised at time $t$. 
Since $\Omega'_{\A,t}=(\omega_{i,\A,t})_{i\in [1,n]}$ can be derived from $\Omega_{\A,t}$, in what follows we do not have to mention $\Omega'_{\A,t}$ explicitly. This representation is flexible because it can accommodate intuitive attack goals, such as: (i) attempting to compromise a fixed set of nodes or computers at time $t$;
(ii) attempting to compromise as many nodes or computers as possible at time $t$;
and (iii) attempting to cumulatively compromise as many nodes or computers as possible at time $t$.

\smallskip

\noindent{\bf Attacker's strategy}.
Inspired by the state-of-the-art industrial characterization of sophisticated cyber attacks, such as the Cyber Kill Chain \cite{hutchins2011intelligence} and Mitre's ATT\&CK \cite{MITREATT87:online}, we propose abstracting them into attacker's strategy. Since strategy is often fixed for $t\in [0,T]$, we specify attacker $\A$'s strategy via $\ell\geq 1$ phases, denoted by $\Gamma_\A = \{\gamma_{\A,1},\ldots,\gamma_{\A,\ell}\}$. For example, we have $\ell=7$ for the Cyber Kill Chain and $\ell =12$ for Mitre's ATT\&CK version 7. Since there are different kinds of strategies, we will demonstrate how to use a specific strategy in our case study.

\smallskip
\noindent{\bf Attacker's capability}.
Given an $\ell$-phase strategy $\Gamma_\A$,
attacker $\A$'s capability at phase $z$, where $z \in [1,\ell]$,
is defined as a set of $m_z$ exploits applicable at phase $z$ and available to $\A$, denoted by $\Delta_{\A,z} = \{\psi_{\A,z,1}, \ldots,\psi_{\A,z,m_z}\}$. 
We define $A$'s capability as  vector  
$\Delta_\A=(\Delta_{\A,z})_{z\in [1,\ell]}$. Since $\A$'s capability depends on exploits, 
we define $\A$'s {\em investment} as {$\AI = \sum_{z=1}^\ell \sum_{q=1}^{m_z} cost_{\A,z,q}$},
where $cost_{\A,z,q}$ is the cost to obtain the $q$-th exploit that is applicable at phase $z$.

\smallskip

\noindent{\bf Attacker's decision-making}. At time $t$, attacker $\A$ uses a decision-making algorithm ${\cal F}_{\A,t}$ to make an attack plan, which specifies the utilization of some of $\A$'s capabilities, denoted by $\Lambda_{\A,t}=(\lambda_{\A,1},\ldots,\lambda_{\A,\ell})$ where 
$\lambda_{\A,z}\in \Delta_{\A,z}$ for $1\leq z\leq \ell$.
This can be denoted by
\begin{equation}
\label{eq:attack-plan}
\lambda_{\A,z}\leftarrow {\cal F}_{\A,t}(\Sigma_{\A,t},\Omega_{\A,t},\Gamma_\A,\Delta_{\A,z}),
\end{equation}
where $\Sigma_{\A,t}=(G_{\A,t}, C_{\A,t}, \Phi_{\A,t}, S_{\A,t})$ is $\A$'s knowledge  at time $t$ as described above, $\Omega_{\A,t}$ is $\A$'s goal at time $t$, $\Gamma_\A$ is $\A$'s strategy, and $\Delta_{\A,z}$ is $\A$'s capability at phase $z$ of the attack strategy.
Note the Eq. \eqref{eq:attack-plan} can be extended to consider multiple {exploits} (rather than a single {exploit} $\lambda_{\A,z}$) that may be used in an appropriate manner (e.g., sequential).

{Putting the description together, we denote attacker $\A=(\A_t)_{t\in [0,T]}$ with $\A_t=(\Sigma_{\A,t},\Omega_{\A,t},\Gamma_{\A},\Delta_\A,{\cal F}_{\A,t})$.}

\subsection{Modeling Defender}

Similar to the attacker (or threat) model, we describe a defender $\D$ via {five} attributes: {\em knowledge} (what $\D$ knows), {\em goal} (what $\D$ aims to achieve), {\em strategy}, {\em capability} (what tools $\D$ possesses), and {\em decision-making} (what algorithms $\D$ uses).

\smallskip

\noindent{\bf Defender's knowledge}.
{We define defender $\D$'s knowledge as vector $\Sigma_\D=(\Sigma_{\D,t})_{t\in [0,T]}$ such that $\Sigma_{{\D},t}=(G_{\D},C_{{\D},t}, \Phi_{{\D},t}, S_{{\D},t})$ is $\D$'s
perception of the ground-truth situation $\Sigma_t=(G, C_t, \Phi_t,S_t)$ at time $t$.} 
In the case of {\em full} knowledge, we have $\Sigma_{{\D},t} = \Sigma_{t}$, meaning the defender knows everything about the ground-truth situation. In the more realistic case of {\em partial} knowledge, the defender only knows: (i) the communication graph $G$, namely $G_{\D} = G$ because the network is managed by the defender; (ii) the network configuration $C_t$, namely $C_{{\D},t}= C_t$ because the defender {\color{black}decides which nodes} run which specific implementations; (iii) some information about the ground-truth vulnerabilities associated with the programs, namely $\Phi_{{\D},t}\subseteq \Phi_t$ because the defender may not know the 0-day ones that are known to the attacker; and (iv) some noisy information about the network's ground-truth cybersecurity state $S_t$ because of the false-positives and/or false-negatives in measuring or inferring cybersecurity states.

\smallskip

\noindent{\bf Defender's goal}.
Corresponding to the program-level vs. computer-level distinction, $\D$'s goal can be defined at two levels. At the {\em program} level of $v\in V$, we define $\D$'s goal as vector $\Omega_\D=(\Omega_{\D,t})_{t\in [0,T]}$ such that $\Omega_{\D,t}=(\omega_{v,\D,t})_{v\in V}$ is $\D$'s goal at time $t$, where $\omega_{v,\D,t} \in [0,1]$ is the tolerable probability that program running at $v \in V$ is compromised at time $t$.
For example, $\omega_{v,\D,t}=0$ means that {a successful attack against $v$} cannot be tolerated; $\omega_{v,\D,t}=0.5$ can be interpreted as that compromise of $v$ for at most 50\% of the time can be tolerated. 
At the {\em computer} level, we define $\D$'s goal as vector $\Omega'_\D=(\Omega'_{\D,t})_{t\in [0,T]}$ such that $\Omega'_{\D,t}=(\omega_{i,\D,t})_{i\in [1,n]}$ is $\D$'s goal at time $t$, where $\omega_{i,\D,t} \in [0,1]$ is the tolerable probability that computer $i$ is compromised at time $t$.
One computer-level goal of particular interest is: $\omega_{i,\D,t}{\leq} 1/3$ for $i\in [1,n]$ and $t\in [0,T]$; it describes cyber defense using Byzantine fault-tolerance techniques to tolerate the compromise of a certain threshold of computers \cite{Lynch96}. Since $\Omega'_\D$ can be derived from $\Omega_\D$, we do not have to mention $\Omega'_\D$ except when we discuss computer-level effectiveness.

\smallskip

\noindent{\bf Defender's strategy}.
We specify defender $\D$'s strategies in employing network diversity as a set $\Gamma_\D = \{\gamma_{\D,1},\ldots,\gamma_{\D,\jmath}\}$. For example, $\gamma_{\D,1}$ represents monoculture software stacks (i.e., the baseline strategy), $\gamma_{\D,2}$ represents static diversity, $\gamma_{\D,3}$ represents proactive diversity with fixed intervals, $\gamma_{\D,4}$ represents reactive-adaptive diversity where the employment is triggered by security alerts, 
and $\gamma_{\D,5}$ represents hybrid diversity (i.e., a combination of proactive diversity and reactive-adaptive diversity). A strategy can be accompanied by {\em some} of the following parameters.
(i) the {\color{black}{\em proportion} of nodes in $V$ (re-)employing diversified implementations} 
(e.g., all or some nodes), denoted by $\eta_{\D,1}$;
(ii) the {\em frequency} at which the diversified implementations will be dynamically re-employed at the nodes, denoted by {$\eta_{\D,2}$};
(iii) the {\em condition} under which diversified implementations are re-employed, denoted by $\eta_{\D,3}$.
As an example showing that not every parameter is relevant to every strategy, we note that the preceding (ii) and (iii) are not relevant to the static diversity strategy; this can be indicated by setting $\eta_{\D,2}={\tt NULL}$ and $\eta_{\D,3}={\tt NULL}$. 

\smallskip

\noindent{\bf Defender's capability}. We define defender $\D$'s capability as the diversified implementations that are available to $\D$. Recall that $\hbar$ different programs running in the network (including both applications and operation systems) and each program may have diversified implementations. $\D$'s capability with respect to program $k$, where $1\leq k \leq \hbar$, is a set of $X_k$ diversified implementations, denoted by $\Delta_{\D,k} = \{\delta_{\D,k,1}, \ldots,\delta_{\D,k,X_k}\}$.
We define $\D$'s capability as the vector of diversified implementations that are available to $\D$, denoted by $\Delta_\D=(\Delta_{\D,k})_{k\in [1,\hbar]}$. Since $\D$'s capability depends on the diversified implementations, we define the defender's investment as $\DI = \sum_{k=1}^\hbar \sum_{w=1}^{X_k} cost_{\D,k,w}$,
where $cost_{\D,k,w}$ is the cost for obtaining the $w$-th diversified implementation of program $k$.

\smallskip

\noindent{\bf Defender's decision-making}. At time $t$, $\D$ uses decision-making algorithm ${\cal F}_{\D,t}$ to make a defense plan {$\Lambda_{\D,t}=(\delta_{\D,t}(v))_{v\in V}$}, which specifies how to employ the diversified implementations of the $\hbar$ programs at nodes $v\in V$ at time $t$. Formally, $\Lambda_{\D,t}$ is the output of the defender's decision-making algorithm ${\cal F}_{\D,t}$ on a number of inputs, denoted by
\begin{equation}
\Lambda_{\D,t}\leftarrow {\cal F}_{\D,t}(\Sigma_{\D,t}, \Omega_{\D,t}, \Gamma_\D, \Delta_\D), 
\end{equation}
where $\Sigma_{\D,t}$ is the defender's perception of the ground-truth situation $\Sigma_t=(G, C_t, \Phi_t,S_t)$ at time $t$, $\Omega_{\D,t}$ is the defender's goal, $\Gamma_\D$ is defender's strategy as described above, and $\Delta_{\D}$ is the defender's capability.

Putting the description together, we denote defender $\D=(\D_t)_{t\in [0,T]}$ with $\D_t=(\Sigma_{\D,t},\Omega_{\D,t},\Gamma_{\D},\Delta_\D,{\cal F}_{D,t})$.

\subsection{Modeling Effect of Dynamic Diversity}

\noindent{\bf Modeling local effect of dynamic diversity at the {\em node} level}. At any point in time, a program (i.e., node $v\in V$) is in one of the following three states: {\em vulnerable}, meaning that the program contains a vulnerability but the vulnerability has not been exploited by the attacker; {\em invulnerable}, meaning that the program contains no vulnerability; and {\em compromised}, either because the program contains a vulnerability that has been exploited, or because the underlying OS is compromised (causing any application program running on top of it to be compromised, no matter whether the program contains vulnerability or not). 

\begin{figure}[!htbp]
\vspace{-0.1cm}
\centering
\includegraphics[width=.45\textwidth]{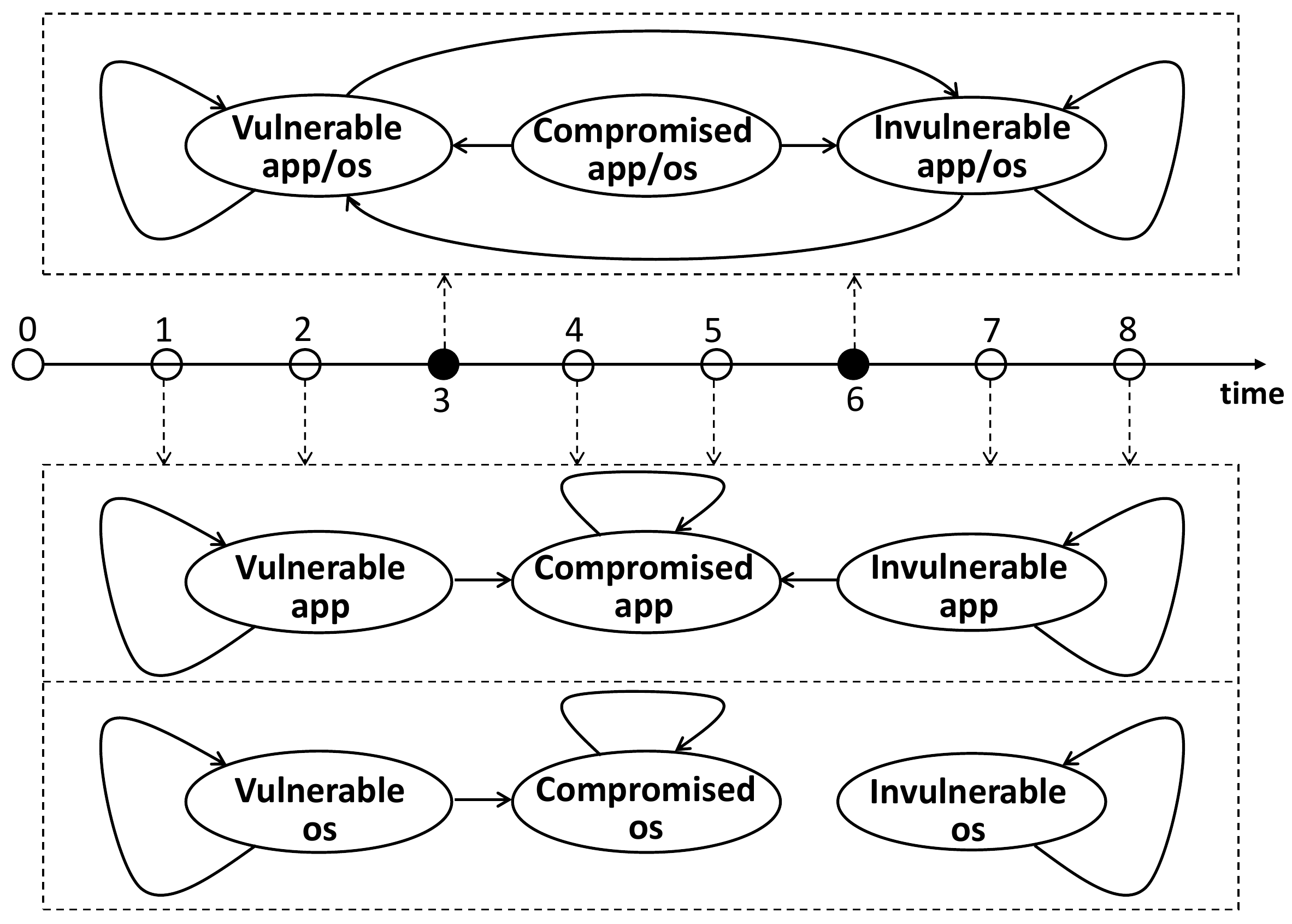}
\caption{Modeling the local effect of dynamic network diversity at the program level (i.e., program running at $v\in V$).
\label{fig:software-replacement}}
\vspace{-0.3cm}
\end{figure}

Figure \ref{fig:software-replacement} highlights the effect of employing dynamic diversity at the program level. Suppose at time $t=3,6$ dynamic diversity is employed such that a different implementation of a program (i.e., application or OS) is employed at time $t$ to replace the implementation employed at time $t-1$. The employment of dynamic diversity at time $t$ leads to one of the following state transitions (see upper half of Figure \ref{fig:software-replacement}): (i) a vulnerable program is replaced with another vulnerable or an invulnerable program of the same functionality; (ii) an invulnerable program is replaced with a vulnerable or another invulnerable program; (iii) a compromised program is replaced with a vulnerable or invulnerable program. However, a vulnerable or invulnerable program is never replaced with a compromised program because a successful attack only occurs to a vulnerable program that has been employed and exploited (as diversified implementations are stored in a secure environment).

At any other point in time than the employment of dynamic diversity (e.g., $t\in \{1,2,4,5,7,8\}$ as shown in Figure \ref{fig:software-replacement}), the security state transition of application programs is different from that of OSes. For applications, the state transitions are: (i) a vulnerable application program stays vulnerable or become compromised, either because its vulnerability is exploited or the underlying OS is compromised; (ii) a compromised application program stays in the compromised state because we focus on the defense based on dynamic diversity, without considering reactive defense that may detect and clean up the {\color{black}compromised application programs}; (iii) an invulnerable application program stays invulnerable or becomes compromised because the underlying OS is compromised. For OSes, the state transitions are: (i) a vulnerable OS stays vulnerable or become compromised because its vulnerability is exploited; (ii) a compromised OS stays in the compromised state because we do not consider reactive defense; (iii) an invulnerable OS stays invulnerable.

\smallskip

\noindent{\bf Quantifying the global effect of dynamic network diversity at the {\em computer} level}. We quantify the global effect at the computer level via the following metrics.

\begin{definition}[time-to-succeed or $\tts$] This metric measures how long it takes attacker $\A$ to break defender $\D$'s goal specified by a program-level vector $\Omega_{\D,t}=(\omega_{v,\D,t})_{v\in V}$ or
{\em computer}-level vector $\Omega'_{\D,t}=(\omega_{i,\D,t})_{i\in [1,n]}$, where $t\in [0, T]$, and $\omega_{v,\D,t}\in [0,1]$ ($\omega_{i,\D,t}\in [0,1]$) is the compromise probability of program $v$ (computer $i$) at time $t$ that can be tolerated.
As mentioned above, a defense goal of particular interest is $\omega_{i,\D,t}{\leq} 1/3 $ for computers $i\in [1,n]$ at any time $t\in[0,T]$ because such compromises can be tolerated by Byzantine fault-tolerant techniques \cite{Lynch96}. Formally, we define $\tts(\A,\D)=\min\{t:\cc(t)> \tau\}$, where $\cc(t)$ is the fraction of compromised computers (i.e., attack damage) at time $t$. 
\end{definition}
The {\em time-to-succeed} metric is reminiscent of the well-known {\em mean-time-to-compromise} metric. However, the former is defined as a random variable with respect to a specific goal (e.g., breaking defender's goal),
where randomness is rooted in different attack strategies, capabilities and decision-makings algorithms. In contrast,  the latter is defined as a number (or the mean value of a random variable). 
\smallskip

\begin{definition}[attacker slow-down or \asd]
Consider attacker $\A$ (i.e., a fixed threat model) that attempts to break defender $\D$'s goal $\Omega'_\D=(\Omega'_{\D,t})_{t\in [0,T]}$, where $\Omega'_{\D,t}=(\omega_{i,\D,t})_{i\in [1,n]}$ and, for concreteness, $\omega_{i,\D,t}{\leq} \tau$ is the tolerable compromise probability for every computer $i\in [1,n]$ and $t \in [0,T]$, while recalling that $\tau=1/3$ corresponds to assuring the assumption that is needed by Byzantine fault-tolerant techniques \cite{Lynch96}.
We define {\em attacker slow-down} as $\asd_{\D,q} = \tts(\A,\D_q) - \tts(\A,\D_1)$, to measure the extent at which $\A$ is slowed down by the employment of network diversity defense strategy $\gamma_{\D,q}$ (denoted by $\D_q$), where $2\leq q \leq5$ in this study, when compared with the baseline strategy $\gamma_{\D,1}$ of monoculture software stacks (denoted by $\D_1$).
\end{definition}

\begin{definition}[attack worst damage or \awd]\label{def:AWD}
This metric measures how much damage an attacker $\A$ can cause in the worst case during $t\in[0,T]$, namely the maximum fraction of compromised computers at any time $t\in[0,T]$. Formally, we define {\em attack worst damage} as $\awd_{G,\hbar,X,Q}(\A,\D, T) = \max\{\cc(t):t\leq T\}$ where 
$G,\hbar,X,Q$ are defined above.
\end{definition}

\begin{definition}[attack extra cost or \aec]
Consider attacker $\A$ (i.e., a fixed threat model) attempting to break defender $\D$'s goal $\Omega'_\D=(\Omega'_{\D,t})_{t\in [0,T]}$, where $\Omega'_{\D,t}=(\omega_{i,\D,t})_{i\in [1,n]}$ and, for concreteness, $\omega_{i,\D,t}{\leq} \tau$ is the tolerable compromise probability for computer $i\in [1,n]$ and $t \in [0,T]$.
We define {\em attack extra cost} ($\aec$) metric, $\aec_{\D,q} = \AI(\gamma_{\D,q}) - \AI(\gamma_{\D,1})$, to measure the number of extra exploits $\A$ needs to obtain (i.e., extra investment) such that $\awd_{G,\hbar,X,Q}(\A,\D_q, T) > \tau$ when $\D$ employs strategy $\gamma_{\D,q}$ (i.e., $\AI(\gamma_{\D,q})$) than the baseline strategy of employing monoculture software stacks (i.e., $\AI(\gamma_{\D,1})$).
\end{definition}
Inspired by the notion of {\em fault-tolerance}, we define a {\em vulnerability-tolerance} metric to capture the upper bound of vulnerabilities in the network that can be tolerated by the defender $\D$ in achieving its goal $\Omega_\D$.
The importance of this metric can be seen as follows. (i) When each computer is vulnerable with probability at least $1/3+{\epsilon}$ for some $\epsilon$, the attacker able to compromise all of them 
can render Byzantine fault-tolerance techniques useless. (ii) When dynamic network diversity is employed, the attacker may only be able to compromise 1/3 of the computers because the  diversity configuration may have changed before the attacker {\color{black}compromises} all of the vulnerable computers. Intuitively, the larger the $\epsilon$, the higher the vulnerability tolerance. 

\begin{definition}[vulnerability-tolerance or \VT] Consider defender goal $\omega_{i,\D,t}{\leq}\tau$ for every computer $i\in [1,n]$ and each time $t\in [0,T]$ and a fixed diversification quality $Q\in [0,1]$ for diversified implementation (when applying the same quality-enhancement techniques), we define $\VT_{\D,q} = \max\{Q:\awd_{G,\hbar,X,Q}(\A,\D_q, T) {\leq} \tau\}$ with $\awd_{G,\hbar,X,Q}(\A,\D_q, T)$  specified in Definition \ref{def:AWD}.
\end{definition}

\begin{definition}[average operational cost or \aoc]
We define the {\em operational cost} incurred by defender's strategy $\gamma_{\D,q}$ at time $t$, denoted by $\oc_{\D,q}(t)\in [0,1]$, as the fraction of programs that are replaced at time $t$, where $2\leq q \leq5$. 
We define the {\em average operational cost} up to time $T$ as $\aoc_{\D,q}(T)=\sum_{t=1}^{T} \oc_{\D,q}(t)/T$, which refers to the average fraction of programs that are re-installed at each time $t\in [1,T]$. Intuitively, the larger the $\aoc_{\D,q}(T)$, the higher the operational cost. Given a defender's goal $\omega_{i,\D,t}\leq\tau$ for computer $i\in [1,n]$ and $t\in [0,T]$, we define $\aoc_{\D,q}^{\sf max} = $ $ \max\{\aoc_{\D,q}(T):\awd_{G,\hbar,X,Q}(\A,\D_q, T)\leq \tau\}$ and $\aoc_{\D,q}^{\sf min} = \min\{\aoc_{\D,q}(T):\awd_{G,\hbar,X,Q}(\A,\D_q, T)\leq \tau\}$ as the maximum and minimum average operational cost to meet the defender's goal, respectively.
\end{definition}

\begin{figure}[!htbp]
\hspace{-0.15cm}
\centering
\includegraphics[width=.5\textwidth, height = .2\textheight]{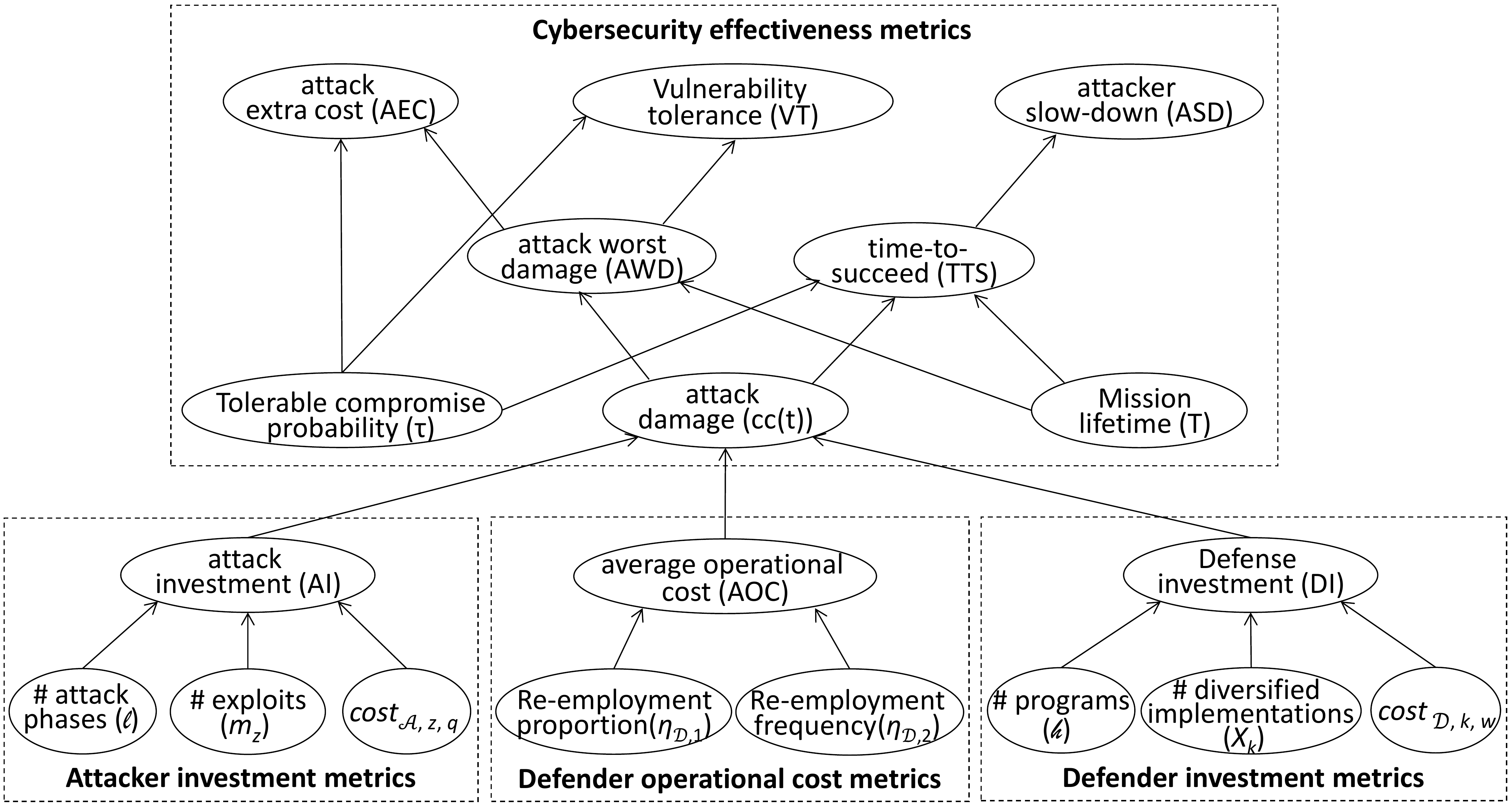}
\caption{{\color{black}Illustration of the relationships between the metrics, where ``$X\to Y$'' means $X$ is a factor in determining $Y$.}
\label{fig:metrics-relations}}
\vspace{-0.3cm}
\end{figure}

Figure \ref{fig:metrics-relations} illustrates the relationship between the metrics. The effectiveness metrics $\tts$, $\asd$, $\awd$, $\aec$ and $\VT$ depend on the attacker's goal in $\tau$, mission lifetime $T$,
and attack damage $\cc(t)$. The $\cc(t)$ depends on the attack investment $\AI$, the defense investment $\DI$, and the defender's operational cost $\aoc$. The attack investment $\AI$ depends on the number of attack phases ($\ell$), the number of exploits applicable at each phase ($m_z$), and the cost to obtain exploits ($cost_{\A,z,q}$). The defense investment depends on the number of programs ($\hbar$), the number of diversified implementations of each program ($X_k$), and the cost to obtain each implementation ($cost_{\D,k,w}$). The defender's average operational cost $\aoc$ depends on the proportion $\eta_{\D,1}$ and frequency $\eta_{\D,2}$ of re-employment.

\subsection{Research Questions (RQs)}
\label{sec:RQs}

In order to characterize the effectiveness of dynamic network diversity, we propose investigating the following RQs, where RQ1-RQ3 correspond to defender's gains, RQ4-RQ5 corresponds to defender's cost.

\begin{itemize}
\item RQ1: To what extent {\color{black}can dynamic network diversity} slow down the attacker?

\item RQ2: How much extra cost can dynamic network diversity impose on the attacker?

\item RQ3: To what extent {\color{black}can dynamic network diversity} increase the defender's vulnerability tolerance? 

\item RQ4: To what extent {\color{black}can dynamic network diversity} increase the defender's average operational cost? 

\item RQ5: Is it true that the more diversified implementations the better? 
\end{itemize}

\section{Simulation Experiments}
\label{sec:case-study}
The preceding framework is meant to be as realistic as we can be. This means that it does not make strong assumptions that would warrant analytical treatment. This explains why we pursue simulation-based empirical study to answer the RQs. We adopt agent-based simulation because agents can conduct activities concurrently, which can mimic real-world attack-defense interaction {\color{black}better than} sequential simulation.
We implement the agent-based simulation via multithreading, where each active local agent, attack and defense alike, is instantiated as one thread such that multiple events can take place concurrently. The simulation experiment is conducted on a computer with 32-CPU and 128GB RAM in the Python environment. In order to accommodate the randomness in the simulation experiment, we conduct 500 simulation runs for each experiment and take their average as the result.
In the discrete-time models, all events occur at discrete points in time. In our experiment, we make every attack or defense activity take effect instantly, which can be extended to accommodate any delays if desired. 

\subsection{Agent-based Simulation of Network System}

\begin{figure*}[!htbp]
\centering
\includegraphics[width=0.85\textwidth]{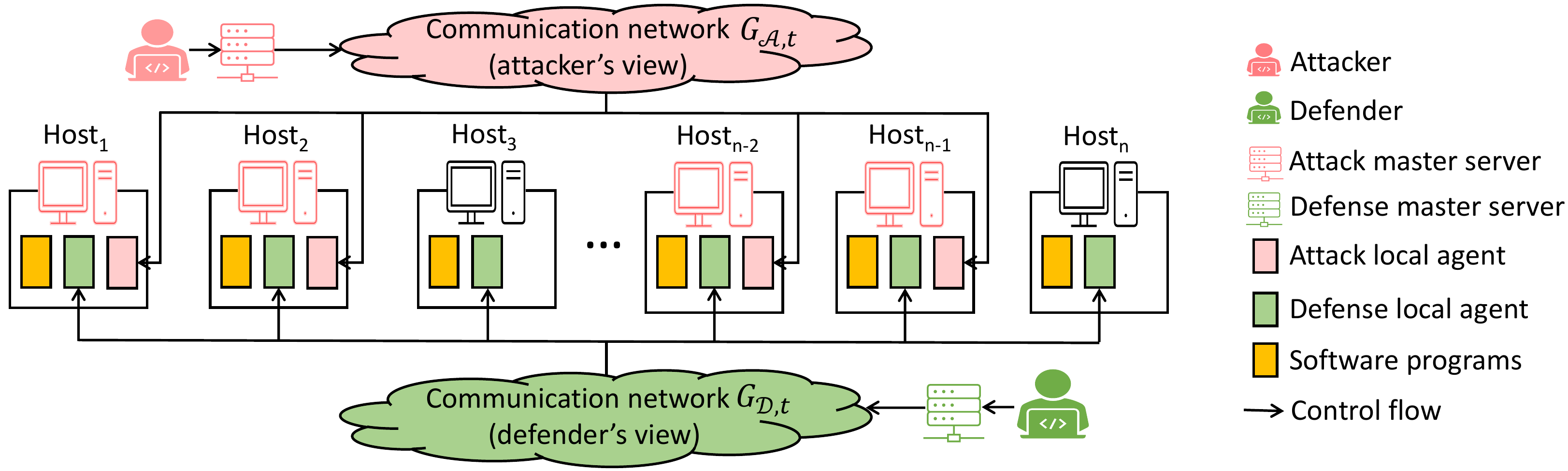}
\caption{Illustration of our agent-based simulation of attack-defense experiments.
\label{multi-agent-architecure}}
\vspace{-0.5cm}
\end{figure*}

Figure \ref{multi-agent-architecure}
illustrates our agent-based simulation of attack-defense experiments. We use a master-agent architecture, where two master agents are responsible for scheduling attacks and defenses, respectively. Each simulated host runs a local defense agent by default, which receives instruction from the defense master server. Once a computer is compromised, a local attack agent is instantiated on the compromised computer to receive instructions from the attack master server. When the employment of dynamic network diversity is conducted {\em manually}, the defense agent can be compromised and re-installed from a clean version when the computer is compromised; when the re-employment process is {\em automatic}, the defense agent must not be compromised (even if the computer is compromised) so as to assure the re-employment of diversified program implementations. The master attack server updates its knowledge $\Sigma_{\A,t}$ based on the information received from the attack agents and makes decisions correspondingly. The master defense server updates its knowledge $\Sigma_{\D,t}$ in a similar fashion.

\smallskip
\noindent{\bf Simulating computers' software stack and communications}.
{\color{black}In order to make the communication graph $G=(V,E)$ as realistic as possible, we adopt two real-world social networks in Twitter and Friendfeed (which is a social media aggregator) \cite{DBLP:conf/asonam/MagnaniR11}, where the former has 5,702 users and the latter has 5,540 users. Together, there are 6,325 users because many users use both Twitter and Friendfeed. We construct the communication graph as follows: a Twitter user corresponds to a Twitter client program and a Friendfeed user corresponds to a Friendfeed client program. A user of Twitter and Friendfeed runs both client programs (i.e., the user's computer is represented as three nodes in the communication graph: one OS and two applications). An edge between two users in the social network means that they communicate with each other using the social network client program. Therefore, the communication graph accommodates the relationships in both social networks.}

\smallskip
\noindent{\bf Simulating network diversity configuration}.
Since there are many ways to configure network diversity and the notion of an {\em optimal} algorithm is elusive (e.g., the algorithms for generating configuration $C_t$ for $t>0$ can be different from the one for generating configuration $C_0$), we will consider multiple algorithms and empirically contrast them. In contrast, monoculture software stack is trivial to configure because each program has exactly one implementation.

\smallskip
\noindent{\bf Simulating software vulnerabilities}.
In order to simulate vulnerabilities contained in the diversified program implementations, namely $\Phi_t$, we assume that each diversified program has the same diversity quality, namely $Q_1 = Q_2 = \ldots =Q_\hbar=Q$, and each implementation of a program is equally vulnerable  with a certain probability; this is a somewhat simplifying assumption but is reasonable in the sense that the same vulnerability prevention and detection techniques may be equally applicable to all implementations. Since there are studies showing that different implementations often do not have the same vulnerability
\cite{han2009effectiveness,garcia2011diversity}, we assume that the vulnerabilities are distinct in the sense that each requiring a different exploit.

\subsection{Simulating Attacker}

\noindent{\bf Simulating attacker's knowledge}.
For describing attacker's knowledge at time $t =0$, we assume that the attacker already compromised some vulnerable programs running at some nodes $v\in V$, namely the initial compromise denoted by $\InitialCompromise$. This is reasonable because initial compromise typically follows reconnaissance or is waged by an insider threat, which is orthogonal to the main purpose of the present study. 
As attack proceeds, the attacker can increase its knowledge by learning more information about the communication graph $G=(V,E)$, and the programs running at the other nodes $v\in V$ that may not be known to the attacker at time $t=0$. We assume the attacker knows the vulnerabilities associated with the diversified programs.

\smallskip

\noindent{\bf Simulating attacker's goal}.
For describing attacker's goal $\Omega_{\A,t}$, we assume the attacker wants to cumulatively compromise as many programs as possible till time $t=T$. That means $\omega_{v,\A,T} = 1$ for $v\in V_{\A,t}$, where $t \in [0,T]$.

\smallskip

\noindent{\bf Simulating attacker's strategy}.
The framework aims to accommodate many attack strategies. In order to make our simulation experiment concrete, we adopt MITRE's ATT\&CK  \cite{MITREATT87:online} because it is widely used. 
The attack {\em tactics} in ATT\&CK can be naturally mapped to the attack phases in our framework. In terms of attacker's strategy $\Gamma_{\A}$, we focus on the following phases: (i) {\em installation}, which corresponds to $\gamma_{\A,1}$ in the framework and 
installing attack agents in compromised computers 
(this phase is called {\em command-and-control} in ATT\&CK);
(ii) {\em discovery}, which corresponds to $\gamma_{\A,2}$ in the framework and allows the attacker to concurrently explore the other programs running at the nodes in the other computers of the network;
(iii) {\em privilege escalation}, which corresponds to $\gamma_{\A,3}$ in the framework and occurs when the attacker gains the root privilege in the compromised computer; (iv) {\em lateral movement}, which corresponds to  $\gamma_{\A,4}$ in the framework and allows the attacker or its malware to exploit the vulnerabilities in the a remote software stack; and (v) {\em causing damages}, which corresponds to $\gamma_{\A,5}$ and allows the attacker to causes damages to a compromised computer (this phase accommodates the tactics of {\em collection}, {\em exfiltration}, and {\em impact} in ATT\&CK).

\smallskip

\noindent{\bf Simulating attacker's capability}.
In order to describe attacker's capability $\Delta_\A$, we adopt ATT\&CK's attack {\em procedures} (i.e., attacks) as exploits. We assume the attacker possesses the following exploits: (i) an exploit for achieving {\em remote access}, denoted by $\psi_{\A,1,1}$;
(ii) an exploit for achieving {\em remote system discovery} or obtaining information about the software running in other computers, denoted by $\psi_{\A,2,1}$;
(iii) an exploit for discovering {\em system information} or getting detailed information about a 
compromised computer, denoted by $\psi_{\A,2,2}$; (iv) a set of exploits for escalating privilege or compromising a vulnerable OS from a compromised application running on top of it, denoted by $\{\psi_{\A,3,1}, \ldots,\psi_{\A,3,m_3}\}$;
(iv) a set of exploits for {\em remote exploitation} capability or compromising a remote, vulnerable computer for lateral movement purposes, denoted by $\{\psi_{\A,4,1}, \ldots,\psi_{\A,4,m_4}\}$; and (v) {an exploit causes some damages (e.g., collecting sensitive data from the compromised software), denoted by 
$\psi_{\A,5,1}$.}
For our purposes, it suffices to assume $cost_{\A,z,q}=1$ for $1\leq z \leq \ell$, $1\leq q\leq {m_z}$, meaning that each exploit incurs the same cost to the attacker (e.g., purchasing or developing an exploit). Future studies can extend this basic scenario to actual cost that may be incurred to the attacker.

\begin{figure}[!htbp]
\centering
\includegraphics[width=.45\textwidth,height=0.06\textheight]{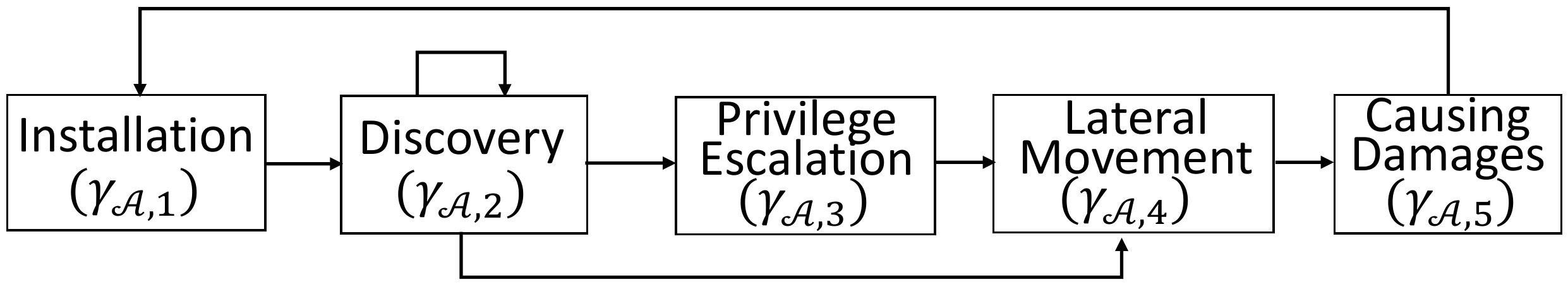}
\caption{Attacker $\A$'s 5-phase strategy (i.e., $\gamma_{\A,1},\ldots,\gamma_{\A,5}$) and possible decisions (i.e., the arrows), which are adopted from ATT\&CK's attack simulator called CALDERA \cite{applebaum2016intelligent}. \label{fig:attacker-strategy}}
\end{figure}

\smallskip

\noindent{\bf Simulating attacker's decision-making}.
In order to simulate the attacker's decision-making function ${\cal F_\A}$, we adopt the decision-making component of ATT\&CK that is used by ATT\&CK's sub-system known as CALDERA \cite{applebaum2016intelligent}.
Figure \ref{fig:attacker-strategy} shows how attacks proceed according to the 5-phase strategy mentioned above: install remote access tools at compromised computers; discover local and remote targets; compromise vulnerable OS via privilege escalation (if applicable); compromise remote computers for lateral movement (if applicable); conduct malicious activities; and repeat these processes. Basically, the decision-making algorithm outputs the next exploit that is to be executed, as follows: (i) if the next phase is $\gamma_{\A,1}$ or $\gamma_{\A,5}$, the attacker will use exploit 
$\psi_{\A,1,1}$ or $\psi_{\A,5,1}$ because it only possesses one exploit at each phase; (ii) if the next phase is $\gamma_{\A,2}$, the attacker will use exploits $\psi_{\A,2,1}$ and $\psi_{\A,2,2}$ simultaneously, where $\psi_{\A,2,1}$ targets one or more remote computers and $\psi_{\A,2,2}$ targets the compromised local computer; (iii) if the next phase is $\gamma_{\A,3}$, the attacker will select one exploit from the set $\{\psi_{\A,3,1}, \ldots,\psi_{\A,3,m_3}\}$ according to the vulnerability information discovered in phase $\gamma_{\A,2}$; (iv) if the next phase is $\gamma_{\A,4}$, the attacker will select one exploit from the set $\{\psi_{\A,4,1}, \ldots,\psi_{\A,4,m_4}\}$ according to the vulnerability information discovered in phase $\gamma_{\A,2}$.

\subsection{Simulating Defender}

\noindent{\bf Simulating defender's knowledge}.
As described in the framework, the defender naturally knows, as a part of its knowledge $\Sigma_{\D,t}$, the communication graph $G$ (i.e., $G_\D$ = $G$) and the network configuration $C_t$ (i.e., $C_{{\D},t}= C_t$) for any past and present time $t$. At time $t=0$, we set that the defender's perception of the cybersecurity state $S_0$ as $S_{\D,0}= (0,0,\ldots,0)_{|V|}$ because the attack-detection tool (if applicable) may start to run at $t=0$.
The defender does not know the information about the vulnerabilities because we allow zero-day vulnerabilities, meaning that $\hat G_{\D,0} = \emptyset$. Putting these together, the defender's initial knowledge is $\Sigma_{\D,0} = (G_\D, C_{\D,0} S_{\D,0}, \hat G_{\D,0})$.     

\smallskip
\noindent{\bf Simulating defender's goal}.
For describing a defender's goal $\Omega_{\D,t}$, we assume the defender aims to keep the compromise rate at any time $t\leq T$ under a certain threshold $\tau_{\D}$. That is, we have $\omega_{i,\D,t}= \tau_{\D}$ for $i\in[1,n]$ and $t\in[0,T]$ on average, where the average is over the 500 simulation runs of each experiment.

\smallskip

\noindent{\bf Simulating defender's strategy}.
We consider the following 5 defense strategies $\Gamma_{\D}$, {(i) {\em monoculture} software stacks, denoted by $\gamma_{\D,1}$, which corresponds to $\eta_{\D,1} = {\tt NULL}$, $\eta_{\D,2}={\tt NULL}$, $\eta_{\D,3}={\tt NULL}$. (ii) {\em Static diversity}, denoted by $\gamma_{\D,2}$, which corresponds to $\eta_{\D,1} = V$, $\eta_{\D,2}={\tt NULL}$, $\eta_{\D,3}={\tt NULL}$. (iii) {\em Proactive diversity}, denoted by $\gamma_{\D,3}$, which corresponds to $\eta_{\D,1}\neq {\tt NULL}$, $\eta_{\D,2}\neq {\tt NULL}$, $\eta_{\D,3}\ = {\tt NULL}$. (iv) {\em Reactive-adaptive diversity}, denoted by $\gamma_{\D,4}$, which corresponds to $\eta_{\D,2} = {\tt NULL}$, $\eta_{\D,3} \neq {\tt NULL}$, because dynamic network diversification is triggered by some security events, which may come from an intrusion detection system 
or the observed network-wide cybersecurity state $S_{{\D},t}$.
Since such reactive intelligence is often noisy in practice,
we incorporate false-negative rate ($\fnr$) and false-positive rate ($\fpr$) into such intelligence. 
When such intelligence is provided by an employed attack detection system, we assume that the attack detection system cannot be compromised in the present study.
(iv) {\em Hybrid diversity}, denoted by $\gamma_{\D,5}$, which combines the aforementioned proactive diversity and reactive-adaptive diversity and corresponds to $\eta_{\D,2} \neq {\tt NULL}$, $\eta_{\D,3} \neq {\tt NULL}$. In this case, dynamic network diversity is triggered periodically or by security events.
}

\smallskip

\noindent{\bf Simulating defender's capability}.
{For simplicity, we assume that  
defender's capability $\Delta_\D$ includes: (i) the same number of diversified implementations for each program (for simplicity), namely $X_1 = X_2 = \ldots =X_\hbar=X$; and (ii) $cost_{\D,k,w}=1$ for $1\leq k \leq \hbar$ and $1\leq w\leq {X_k}$, meaning that each diversified implementation incurs the same cost to the defender. These simplifying assumptions, while arguably reasonable when diversification is an automated process, need to be extended to consider more general cases. It is worth mentioning that different versions of a program should not be counted as diversified implementations because they would have many vulnerabilities in common.

\smallskip

\noindent{\bf Simulating defender's decision-making}.
For describing defender's decision-making function ${\cal F}_{\D,t}$, we first consider ${\cal F}_{\D,0}$ at time $t$ = 0, which outputs the initial network diversity configuration $C_0$. We consider three decision-making algorithms for ${\cal F}_{\D,0}$ at time $t=0$}: (i) the baseline {\em random coloring} algorithm, which assigns a random implementation of the program in question to run at node $v$;
(ii) the {\em color flipping} algorithm \cite{ODonnellCCS2004}, which uses the random coloring algorithm mentioned above as a starting point and then iteratively let nodes change their colors to reduce the number of {\em defective} edges (i.e., the edges with two end nodes having the same color or running the same implementation of a software);
(iii) a new algorithm we propose, which leverages the degrees of the nodes to assign software implementations to nodes $v\in V$, by giving the large-degree nodes a high priority in running diversified programs. Our algorithm is different from the {\em color flipping} algorithm mentioned above because we prioritize large-degree nodes in the initial assignment. Due to space limit, we defer the pseudo-code of the algorithm to the Supplementary Material.

For describing defender's decision-making function ${\cal F}_{\D,t}$ for $t>0$, we consider ${\cal F}_{\D,1} = \ldots={\cal F}_{\D,T} = {\tt NULL}$ for defense strategy $\gamma_{\D,2}$ because the software deployment stays unchanged over time in the case of static network diversity. For dynamic diversity $\gamma_{\D,q}$ where $3\leq q \leq5$, we consider a simple random decision-making function ${\cal F}_{\D,t}$ for $t\in [1, T]$, namely that the defender randomly selects diversified implementations to replace the currently-employed implementations at some or all of the nodes $v\in V$ according to defender's strategy. More sophisticated decision-making functions are left for future studies.

\subsection{Simulating Local Effect and Quantifying Global Effect of Dynamic Diversity}
We simulate the local effect of dynamic network diversity at each node $v\in V$ as described in Figure \ref{fig:software-replacement} and the global effect according to the COODAL based attack-defense interactions described in Figure \ref{fig:diversity-model}. We collect the network-wide cybersecurity state and situation over time $t$ to quantitatively answer the RQs (Section \ref{sec:RQs}).

\begin{figure*}[htbp]
\centering
\hspace{-0.5cm}
\subfloat[$\asd_{\D,2}$ w/ static diversity \label{fig:rq1-1}]{\includegraphics[width=.26\linewidth]{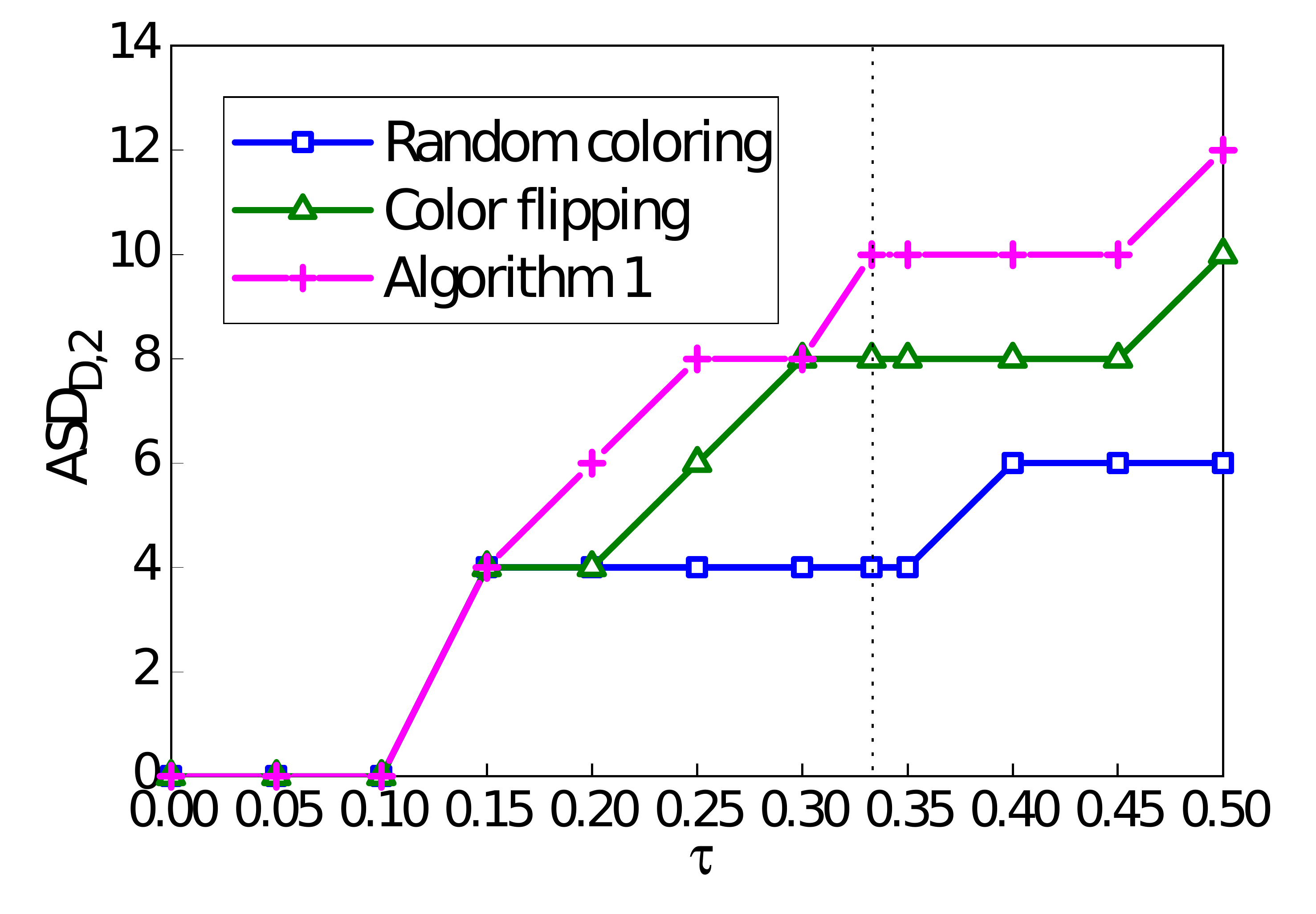}}
\hspace{-0.2cm}
\subfloat[$\asd_{\D,3}$ w/ proactive diversity \label{fig:rq1-2}]{\includegraphics[width=.26\linewidth]{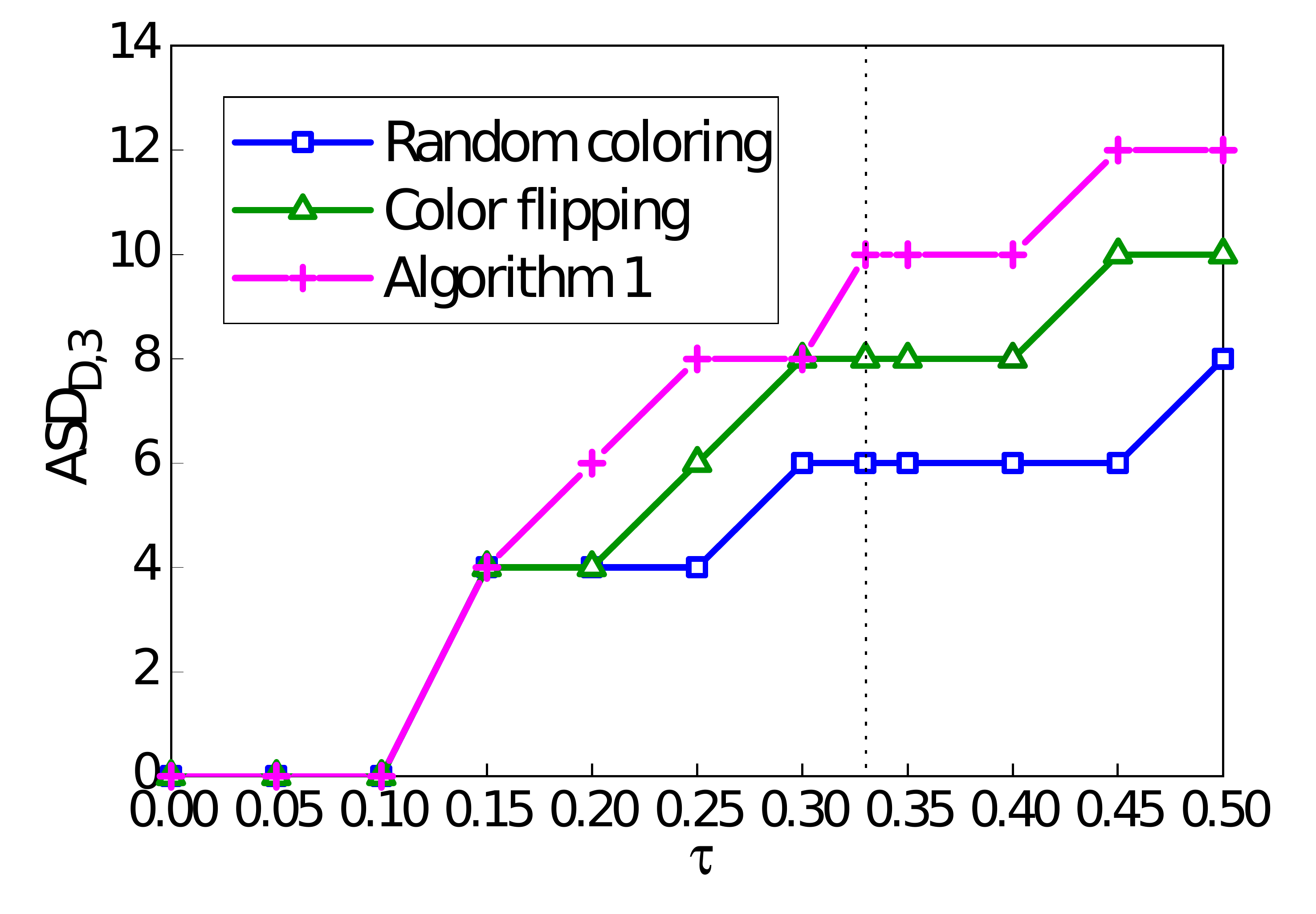}} 
\hspace{-0.2cm}
\subfloat[$\asd_{\D,4}$ w/ reactive diversity \label{fig:rq1-3}]{\includegraphics[width=.26\linewidth]{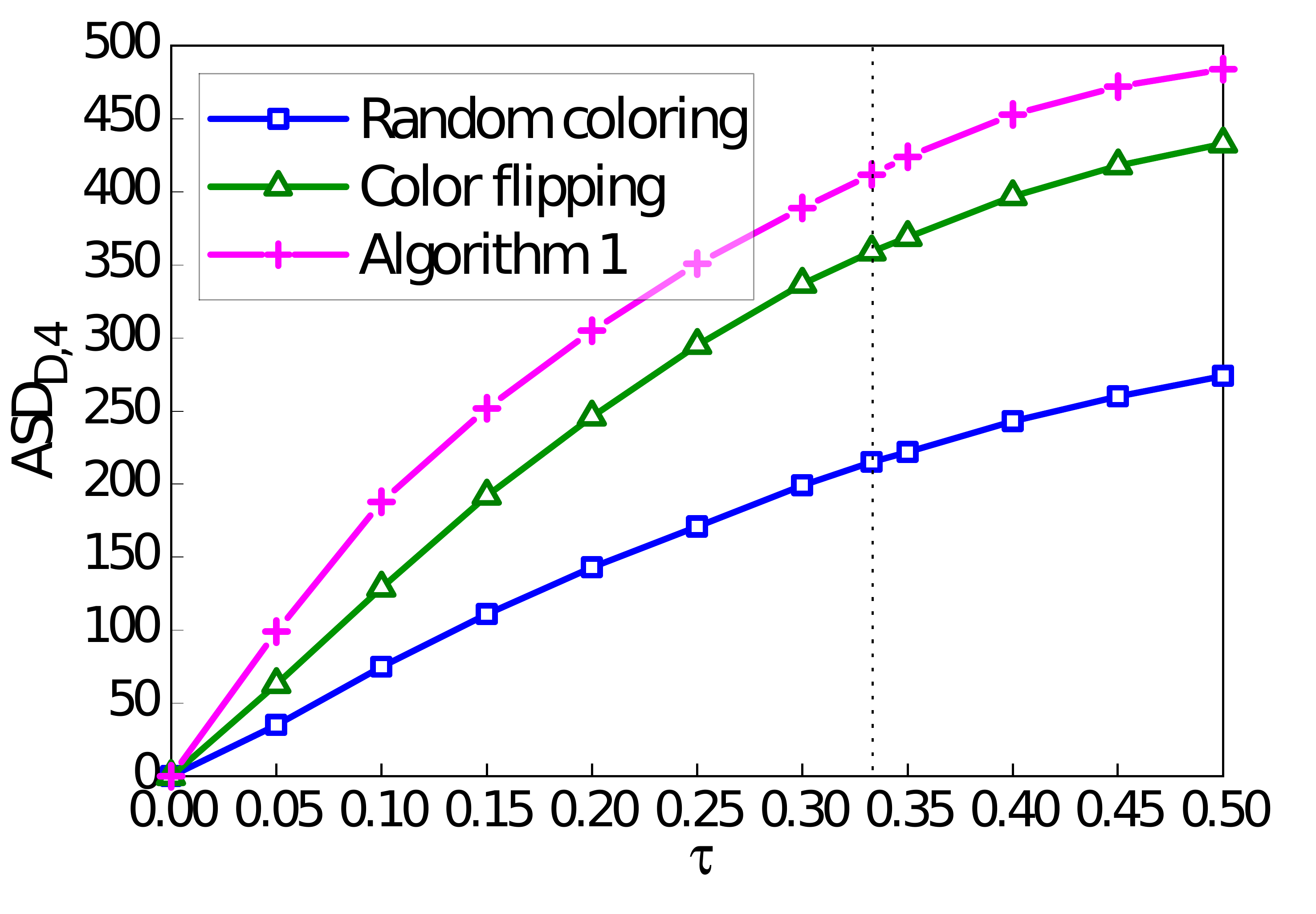}}
\hspace{-0.2cm}
\subfloat[$\asd_{\D,5}$ w/ hybrid diversity \label{fig:rq1-4}]{\includegraphics[width=.26\linewidth]{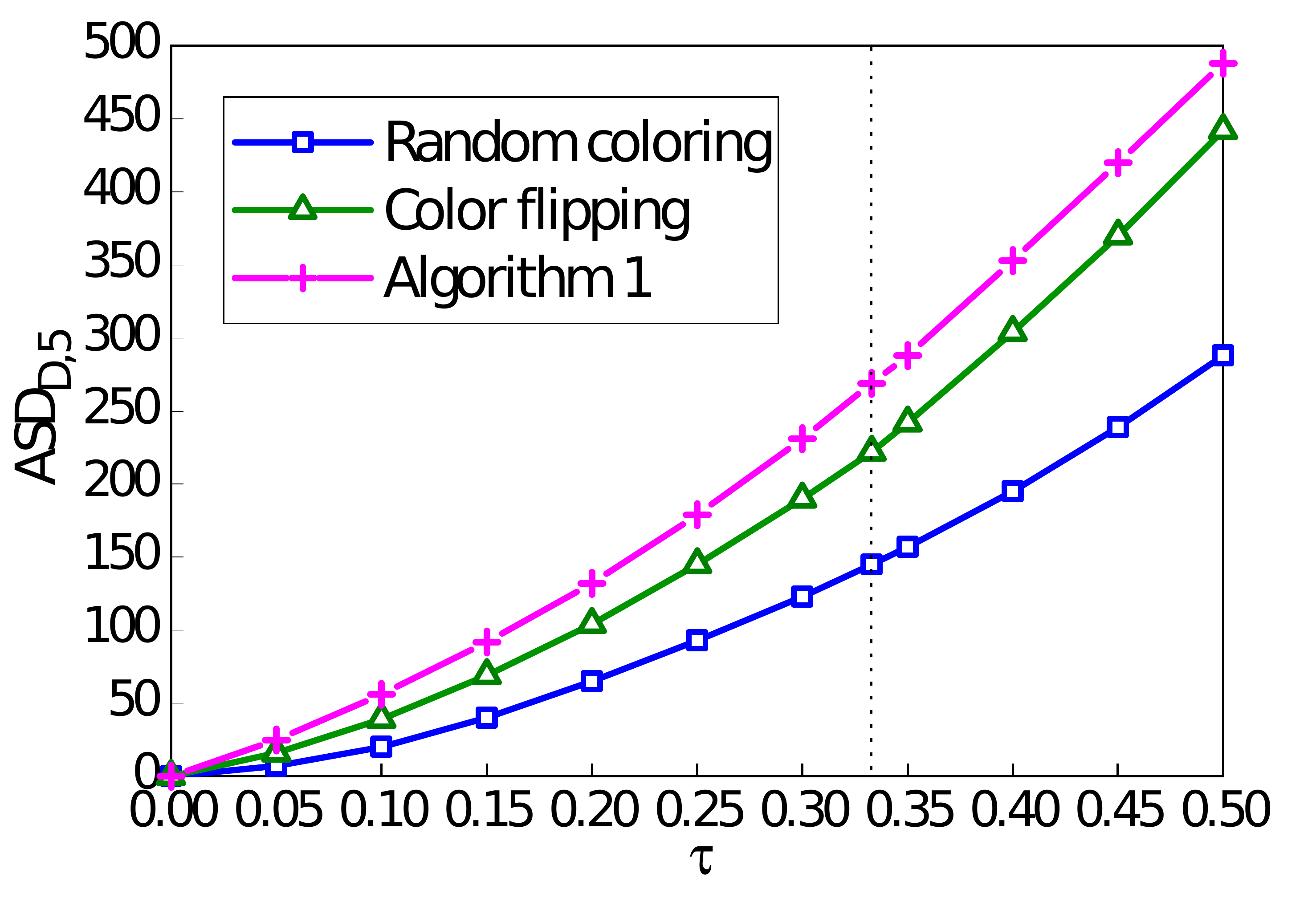}}
\caption{Plots of attacker slow-down $\asd_{\D,q}$ under different diversity strategies (where reactive is short for reactive-adaptive) with varying tolerable compromise threshold $\tau$, where $2\leq q \leq5$ and dotted vertical lines indicate $\tau$ = 1/3. \label{fig:rq1}}
\vspace{-0.3cm}
\end{figure*}

\subsection{Answering RQs}
Our simulation study is centered at measuring the metrics to quantify the security effectiveness of employing network diversity, and then leveraging these quantitative effectiveness to draw insights. {\color{black}While the model is general, the simulation study can only consider some specific parameter settings because it is not feasible to consider all parameter settings. That is, the simulation study only corresponds to some scenarios of the general model, and the findings drawn from the simulation study may not be generalized to other parameter settings. 
Researchers and practitioners can apply our model to their specific parameter settings.
}

\subsubsection{RQ1: To what extent {\color{black}can dynamic network diversity} slow down the attacker?}
In order to answer RQ1, we investigate how the attacker slower-down metric $\asd_{\D,q}$ depends on defender's goal $\tau$ with $2\leq q \leq5$. The experimental parameters are: $\hbar$ = 3 (3 different programs running in the network: Twitter, Friendfeed, OS), $X$ = 10 (each program has 10 diversified implementations), $Q$ = 1 (every diversified implementation is vulnerable), $|\InitialCompromise|$ = 10 (10 programs/nodes are initially compromised), $m_3$ = 5 (the attacker has 5 exploits against OS), $m_4$ = 10 (the attacker has 5 exploits against Twitter and 5 exploits against Friendfeed), and $T$ = 500 (the simulation stops at $t$ = 500).

Figure \subref*{fig:rq1-1} plots attacker slow-down $\asd_{\D,2}$ with respect to defender's goal $\tau\in [0, 0.5]$, where the defender uses static diversity and different decision-making algorithm ${\cal F}_{\D,0}$ at time $t=0$. We observe: (i) $\asd_{\D,2}$ = 0 when $\tau \leq 0.1$, meaning that {static diversity cannot slow down the attacker when the defender can only tolerate no more than 10\% {\color{black}of the nodes being compromised}.} (ii) $\asd_{\D,2} >0 $ when $\tau \geq 0.15$, 
meaning that static diversity can slow down the attacker at an extent that increases with the degree of tolerable compromise. This is reasonable because the attacker has to do more lateral movements in order to disrupt the defender's goal.
(iii) For a fixed tolerable compromise threshold $\tau$, the (initial diversity) decision-making algorithm ${\cal F}_{\D,0}$ matters and our algorithm slows down the attacker most, with an average slow-down that is almost 2X of that of the random coloring algorithm, where the average is over the $\tau$'s.

Figure \subref*{fig:rq1-2} plots attacker slow-down $\asd_{\D,3}$ with respect to defender's goal $\tau\in [0, 0.5]$, where the defender uses different decision-making algorithms ${\cal F}_{\D,0}$ for initial diversity and proactively uses ${\cal F}_{\D,t}$ for $t>0$ with parameters $\eta_{\D,1}$ = 0.5 and $\eta_{\D,2}$ = 0.2 (diversified implementations are re-deployed at 50\% of the nodes every 5 time steps). We make the same observations as that of $\asd_{\D,2}$. This is reasonable because proactive diversity can replace compromised programs with secure programs (i.e., benefiting the defender), but can  also replace secure programs with vulnerable ones (i.e., benefiting the attacker).

Figure \subref*{fig:rq1-3} plots attacker slow-down $\asd_{\D,4}$ with respect to defender's goal $\tau\in [0, 0.5]$, where the defender uses some decision-making algorithm ${\cal F}_{\D,0}$ for initial diversity and reactive-adaptively uses ${\cal F}_{\D,t}$ for $t>0$ while assuming parameters $\fpr$ = 0.1 and $\fnr$ = 0.1 (i.e., the attack-detection or threat intelligence has a 10\% false-positive rate and a 10\% false-negative rate).
We make the following observations: (i) $\asd_{\D,4} >0 $ for $\tau \in (0, 0.5]$, meaning that reactive-adaptive diversity can always 
slow down the attacker at an extent that concavely increases with the tolerable compromise threshold $\tau$.
This is reasonable because the defender can replace a likely-compromised software with another diversified implementation to benefit the defender.
(ii) For a fixed tolerable compromise threshold $\tau$, the initial diversity algorithm ${\cal F}_{\D,0}$ matters because our algorithm slows down the attacker most, with an average of almost 2X slow-down than that of the the random coloring algorithm.

Figure \subref*{fig:rq1-4} plots attacker slow-down $\asd_{\D,5}$ with respect to defender's goal $\tau\in [0, 0.5]$, where the defender uses some decision-making algorithm ${\cal F}_{\D,0}$ for initial diversity and hybrid {(of proactive and reactive-adaptive)} decision-making algorithm 
${\cal F}_{\D,t}$ for $t>0$ with parameters $\eta_{\D,2}$ = 0.2, $\fpr$ = 0.1 and $\fnr$ = 0.1.
We observe the same phenomena as in the case of reactive-adaptive diversity, except that the degree of slow-down incurred by  hybrid diversity increases with the tolerable compromise threshold in a {\em convex} (rather than {\em concave}) fashion.

{\color{black}
By comparing Figures \subref*{fig:rq1-1}-\subref*{fig:rq1-4}}, we observe that for a fixed (initial diversity) decision-making algorithm ${\cal F}_{\D,0}$, we have $\asd_{\D,4} > \asd_{\D,5} \gg \asd_{\D,3} \approx \asd_{\D,2} $ for $\tau \in (0, 0.45]$, indicating that reactive-adaptive diversity outperforms hybrid diversity, which significantly outperforms proactive diversity and static diversity.
Consider $\tau = 1/3$ as an example, we observe $\asd_{\D,4}$ = 412, $\asd_{\D,5}$ = 269, $\asd_{\D,3}$ = 10, and $\asd_{\D,2}$ = 10 when using our algorithm as (initial diversity) decision-making algorithm ${\cal F}_{\D,0}$. 
{\color{black}Note that the curves in Fig. \subref*{fig:rq1-1} are not smooth because, under the static diversity strategy, compromised programs are not cleaned up and can attack others. This also {\color{black}explains} why the curves in Fig. \subref*{fig:rq1-2} are not smooth, namely that proactive diversity always periodically selects random programs for re-employing dynamic diversity, {\color{black}causing some compromised computers to remain compromised for extended durations and allowing them to conduct further attacks.} 
In contrast, the curves in Fig. \subref*{fig:rq1-3} and Fig. \subref*{fig:rq1-4} are smooth because these two strategies leverage reactive defense systems to identify and replace the compromised programs timely, which can limit the abrupt spreading of attacks and slow down the attacker. One reason for the reactive-adaptive strategy to perform better than the hybrid strategy is that the former can immediately clean up the compromised programs, but the latter waits until a period of time, which allows the attacker to compromise other vulnerable programs. That is, the difference is caused by whether there is a gap between when compromised programs are detected and when compromised programs are cleaned up.}

\begin{insight}
\label{insight:RQ1}
In terms of the attacker slow-down metric, reactive-adaptive diversity is the most effective strategy and the initial diversity configuration matters.
\end{insight}

\ignore{
These lead to: 
\begin{insight}
Given a defender that can tolerate no more than 50\% compromises, reactive-adaptive diversity significantly improves attacker slow-down, followed by hybrid diversity, when compared with proactive diversity and static diversity.
\end{insight}
}

\subsubsection{RQ2: How much extra cost can dynamic network diversity impose on the attacker?}
In order to answer RQ2, we investigate how the {\em attack extra cost} metric $\aec_{\D,q}$ increases with defender's goal $\tau$, where $2\leq q \leq5$. For this purpose, we need to see how attack cost affects the network-wide cybersecurity state, {especially attack worst damage $\awd_{G,\hbar,X,Q}(\A,\D_q, T)$, where $2\leq q \leq5$}. The simulation experiment parameters are: $\hbar = 3$ (i.e., 3 different programs running in the network: Twitter, Friendfeed, and OS), $X = 10$ (i.e., each program has 10 diversified implementations), $Q = 1$ (i.e., every diversified implementation is vulnerable), $|\InitialCompromise| = 10$ (i.e., 10 programs or nodes are initially compromised), ${\cal F}_{\D,0}$ is our algorithm for employing initial diversity, $\eta_{\D,1} = 0.5$ (i.e., diversified implementations are dynamically re-employed at 50\% of all nodes), $\eta_{\D,2} = 0.2$ (i.e., diversified implementations are re-employed every 5 time steps), $\fpr = 0.1$ (i.e., 10\% false-positive rate in detecting attacks), $\fnr = 0.1$ (i.e., 10\% false-negative rate in attack detection), $T = 500$. We assume that the attacker uses the available exploits together, for the sake of reducing the uncertainty in the outcomes that may be incurred by the orders of exploits usage. 

\begin{figure}[!htbp]
\vspace{-0.1cm}
\centering
\hspace{-0.5cm}
\subfloat[$\awd_{G,\hbar,X,Q}(\A,\D_q, 500)$ \label{fig:rq2-1}]{\includegraphics[width=.54\linewidth]{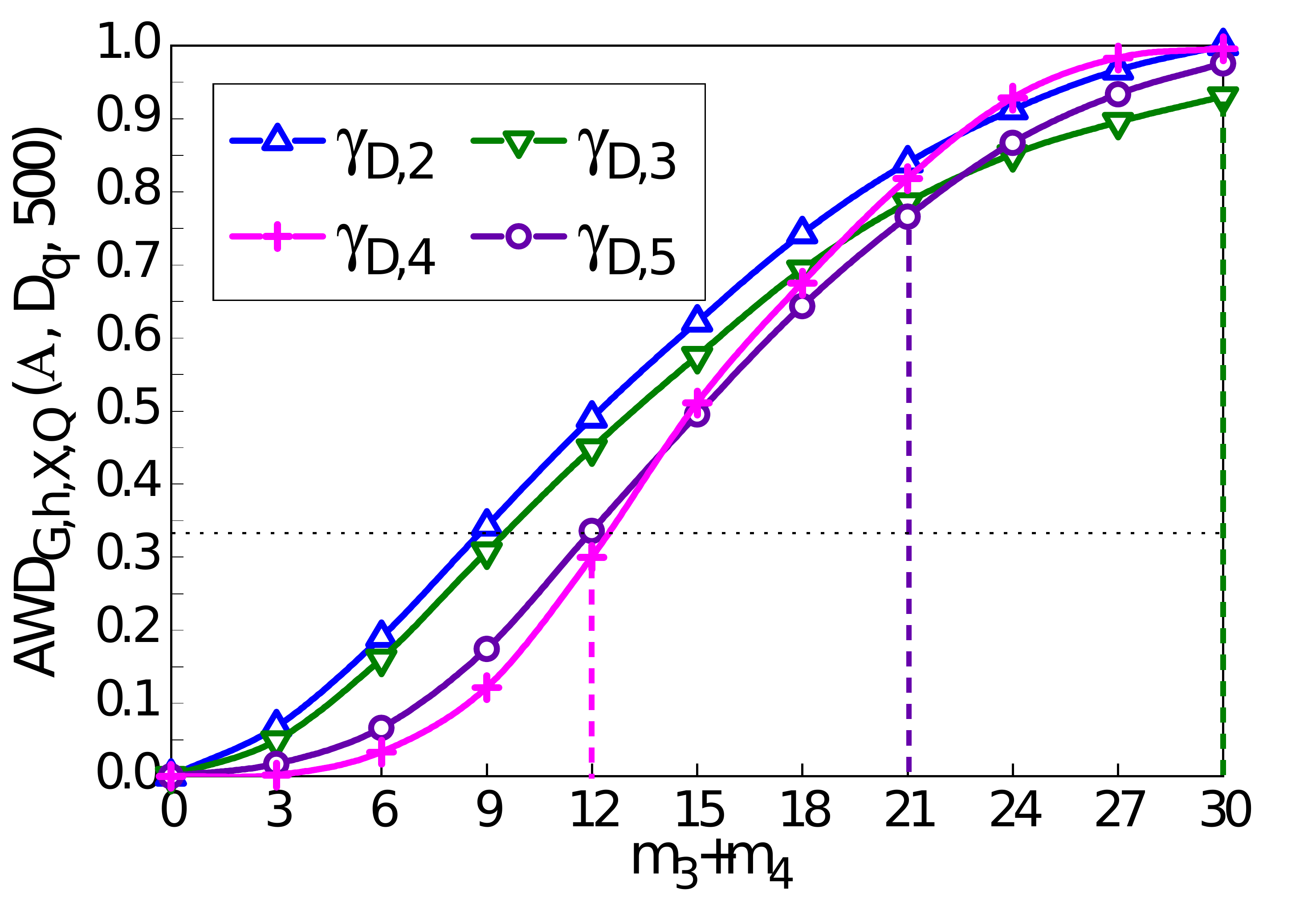}}
\hspace{-0.4cm}
\subfloat[$\aec_{\D,q}$ \label{fig:rq2-2}]{\includegraphics[width=.54\linewidth]{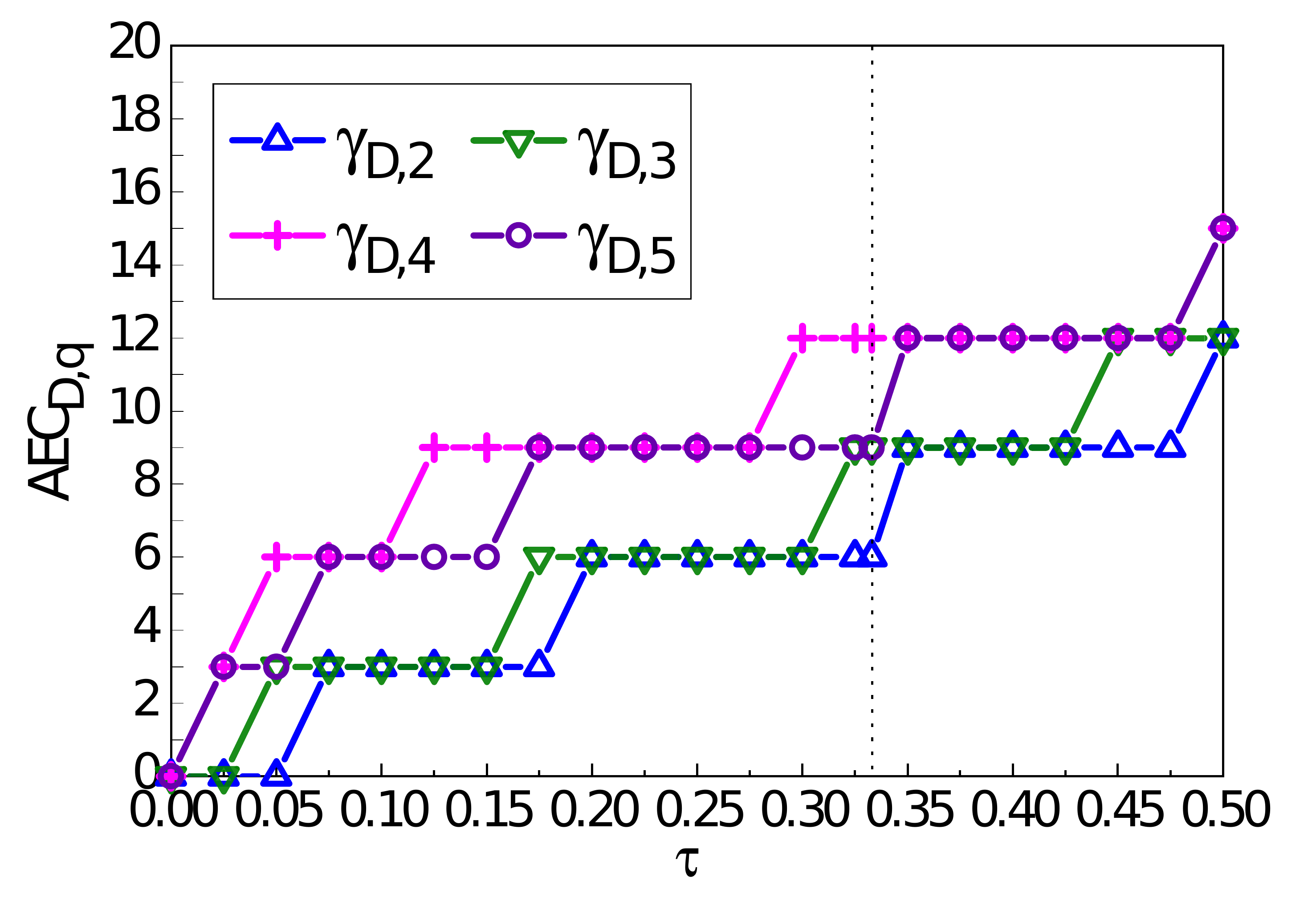}}
\hspace{-0.3cm}
\caption{\label{fig:rq2} Plots of $\awd_{G,\hbar,X,Q}(\A,\D_q, 500)$ and $\aec_{\D,q}$ with $2\leq q \leq5$.
}
\vspace{-0.1cm}
\end{figure}

{Figure \subref*{fig:rq2-1} plots $\awd_{G,\hbar,X,Q}(\A,\D_q, 500)$ for $2 \leq q  \leq 5$,} with respect to the total number $m_3 + m_4$ of exploits possessed by the attacker,  where $m_3$ is the number of exploits that provide the attacker with privilege escalation capability and $m_4$ is the number of exploits that provide the attacker with lateral movement capability. We observe the following {\em phase-transition} phenomenon. When $0<m_3+m_4\leq 12$ (i.e., the attacker possessing no more than 40\% of the total 30 exploits, which correspond to all of the 30 vulnerabilities in the diversified implementations), the reactive-adaptive diversity strategy $\gamma_{\D,4}$ leads to the lowest attack worst damage (i.e., the $\gamma_{\D,4}$ curve); when $12 < m_3+m_4\leq 21$, the hybrid diversity strategy $\gamma_{\D,5}$ leads to the lowest attack worst damage  (i.e., the $\gamma_{\D,5}$ curve); when $21 < m_3+m_4\leq 30$, the proactive diversity strategy $\gamma_{\D,3}$ leads to the lowest attack worst damage  (i.e., the $\gamma_{\D,3}$ curve). {\color{black}This phenomenon can be explained as follows. In terms of the {\em attack worst damage} metric, 
reactive-adaptive diversity is the most effective strategy against a less capable attacker because the defender can detect and replace the small number of compromised programs; proactive diversity is the most effective strategy against a more capable attacker, which compromises a large number of computers and demands periodic enforcement of dynamic diversity at most, if not all, of the computers.}

From Figure \subref*{fig:rq2-1}, we can derive the attack extra cost $\aec_{\D,q}$ with respect to the defender's goal $\tau\in[0,0.5]$ {(i.e., no more than a $\tau$ fraction of the nodes are compromised at {\em any} time $t\in[0,T]$)} where $2\leq q \leq5$, which is plotted in Figure \subref*{fig:rq2-2}. We make the following observations: (i) $\aec_{\D,4}>0$ and $\aec_{\D,5}>0$ when $\tau \geq 0.025$, $\aec_{\D,3}>0$ when $\tau \geq 0.05$, and $\aec_{\D,2}>0$ when $\tau \geq 0.075$, meaning that employing dynamic network diversity defense strategy can impose attack extra cost on the attacker at an extent that increases with the degree of tolerable compromises. 
(ii) $\aec_{\D,4} \geq \aec_{\D,5} \geq\aec_{\D,3} \geq\aec_{\D,2}$ always holds for any $\tau\in(0,0.5]$. Consider $\tau$ = 1/3 as an example, we observe $\aec_{\D,4} = 40\% \geq \aec_{\D,5} = 30\% \geq\aec_{\D,3} = 30\% \geq\aec_{\D,2} =20\%$, meaning that reactive-adaptive diversity imposes extra cost on the attacker more than hybrid diversity, which incurs more than proactive diversity and even more than static diversity.

\begin{insight}
\label{insight:RQ2}
In order to reduce the attack worst damage, different diversity strategies should be used in different parameter regimes.
\end{insight}

\subsubsection{RQ3: To what extent {\color{black}can dynamic network diversity} increase the defender's vulnerability tolerance? }
In order to answer RQ3, we investigate how the {\em vulnerability tolerance} metric $\VT_{\D,q}$ depends on defender's goal $\tau$, where $2\leq q \leq5$. {Since this dependence would rely on attack worst damage $\awd_{G,\hbar,X,Q}(\A,\D_q, T)$ and the parameters $G$ and $\hbar$ are largely determined by the applications and $X$ is largely determined by what is available, we will investigate the impact of diversification quality $Q$.} The simulation experiment parameters are: $\hbar = 3$ (3 different programs running in the network: Twitter, Friendfeed, and OS), $X = 20$ (each program has 20 diversified implementations), $|\InitialCompromise| = 10$ (10 programs or nodes are initially compromised), ${\cal F}_{\D,0}$ is our algorithm for employing initial diversity, $\overline{m_3} = 0.5\times X \times Q$ (the attacker has 50\% of the total $X\times Q$ exploits against the OSes on average), $\overline{m_4} = 0.5\times 2X \times Q$ (the attacker has 50\% of the total $X\times Q$ exploits against Twitter and 50\% of the total $X\times Q$ exploits against Friendfeed on average), $\eta_{\D,1} = 0.5$ (diversified implementations are dynamically re-employed at 50\% of all nodes), $\eta_{\D,2} = 0.2$ (diversified implementations are re-employed every 5 time steps), $\fpr = 0.1$ (10\% false-positive rate in detecting attacks), $\fnr = 0.1$ (10\% false-negative rate in attack detection), and mission lifetime $T = 500$.

\begin{figure}[!htbp]
\vspace{-0.1cm}
\centering
\hspace{-0.5cm}
\subfloat[$\awd_{G,\hbar,X,Q}(\A,\D_q, 500)$ \label{fig:rq3-1}]{\includegraphics[width=.54\linewidth]{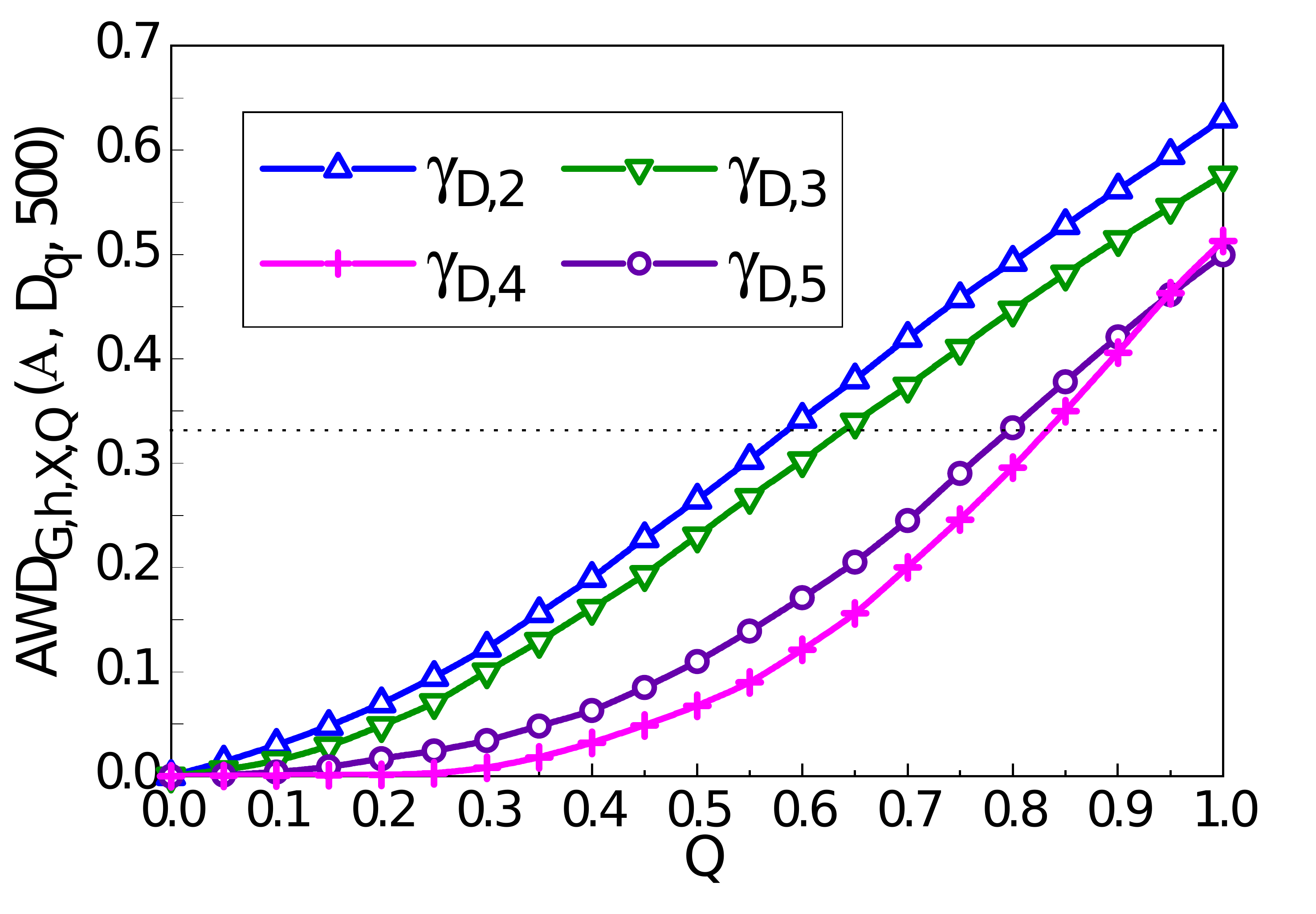}}
\hspace{-0.4cm}
\subfloat[$\VT_{\D,q}$ \label{fig:rq3-2}]{\includegraphics[width=.54\linewidth]{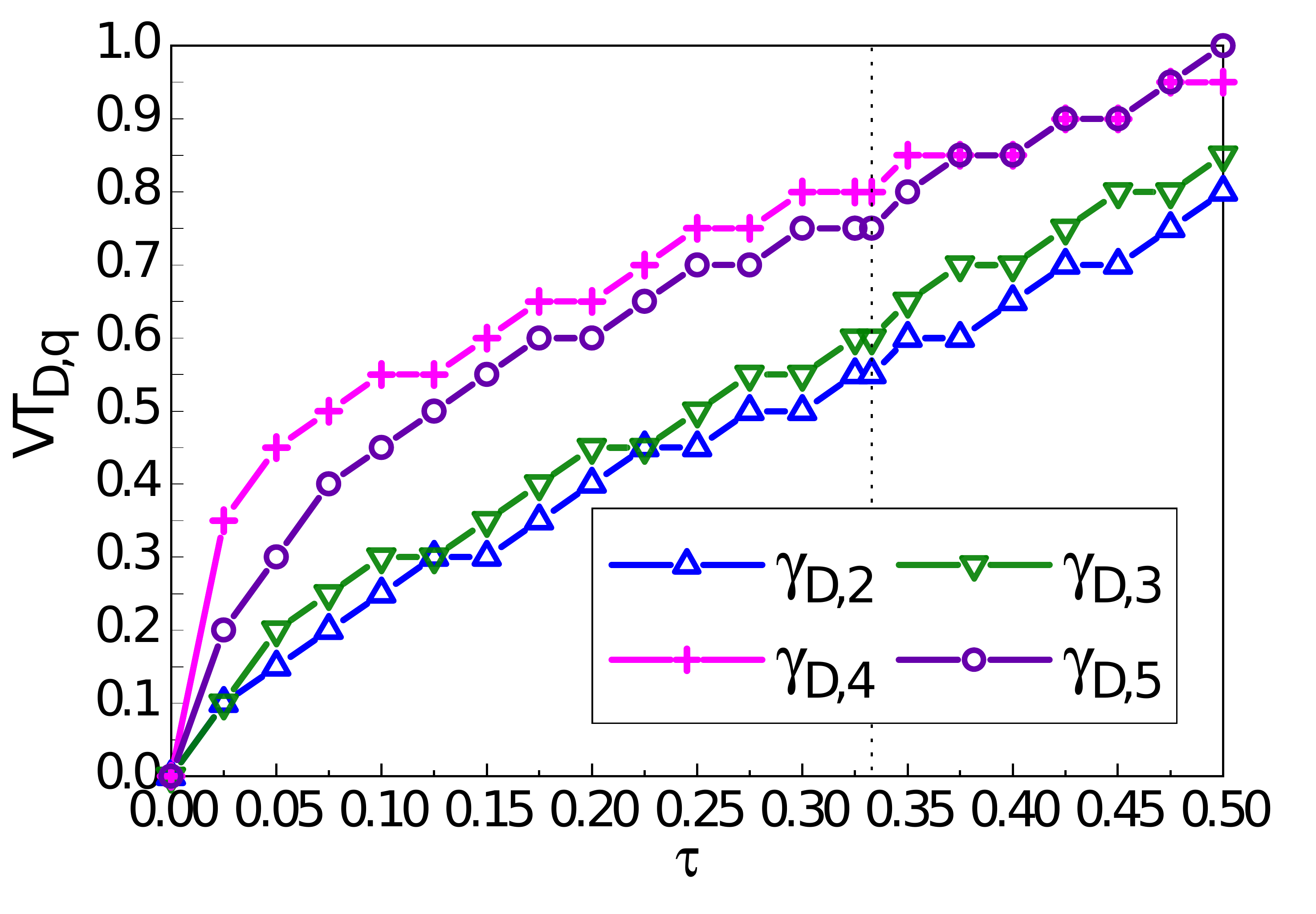}}
\hspace{-0.3cm}
\caption{\label{fig:rq3} Plots of $\awd_{G,\hbar,X,Q}(\A,\D_q, 500)$ and $\VT_{\D,q}$ with $2\leq q \leq5$.}
\vspace{-0.1cm}
\end{figure}

Figure \subref*{fig:rq3-1} plots $\awd_{G,\hbar,X,Q}(\A,\D_q, 500)$ with respect to diversity quality $Q$ when the defender employs network diversity defense strategy $\gamma_{\D,q}$, where $2\leq q \leq5$. We observe that $\cc_{\D,4}< \cc_{\D,5}<\cc_{\D,3}<\cc_{\D,2}$ when $0<Q\leq0.95$, and $\cc_{\D,5}< \cc_{\D,4}<\cc_{\D,3}<\cc_{\D,2}$ when $0.95\leq Q\leq1$. This means that employing dynamic network diversity always leads to higher security than static diversity regardless of the diversification quality $Q$. {We observe that the attack worst damage $\awd_{G,\hbar,X,Q}(\A,\D_q, 500)$ increases when the quality of diversified implementations drops, where $2\leq q \leq5$. This means that dynamic network diversity leads to an even higher security when the diversity quality is high (i.e., low $Q$'s).}

From Figure \subref*{fig:rq3-1}, we can derive the  vulnerability tolerance $\VT_{\D,q}$ with respect to the defender's goal $\tau\in[0,0.5]$ where $2\leq q \leq5$, which is plotted in Figure \subref*{fig:rq3-2}. We observe that $\VT_{\D,4}\geq \VT_{\D,5}>\VT_{\D,3}\geq\VT_{\D,2}$ when $0<\tau\leq0.475$. Consider $\tau$ =1/3 as an example, we observe $\VT_{\D,4}$ = 0.8, $\VT_{\D,5}$ = 0.75, $\VT_{\D,3}$ = 0.6, and $\VT_{\D,2}$ = 0.55. This means that given an attacker that can exploit 50\% of the vulnerabilities in the network on average, reactive-adaptive diversity strategy $\gamma_{\D,4}$ makes the defender tolerate a $0.8-2/3=0.14$ or 14\% extra vulnerabilities, and hybrid diversity strategy $\gamma_{\D,5}$ make the defender tolerate a $0.75-2/3=0.09$ or 9\% extra vulnerabilities; however, proactive and static diversity strategies cannot achieve this effectiveness.

\begin{insight}
\label{insight:RQ3}
Reactive-adaptive diversity leads to a higher vulnerability-tolerance than proactive diversity does.
\end{insight}

\subsubsection{RQ4: To what extent {\color{black}can dynamic network diversity} increase the defender's average operational cost?}
In order to answer RQ4, we investigate how the {\em average operational cost} metric $\aoc_{\D,q}$ increases with defender's goal $\tau$, where $2\leq q \leq5$. For this purpose, we need to know how attack worst damage $\awd_{G,\hbar,X,Q}(\A,\D_q, T)$ depends on defense diversity strategies $\gamma_{\D,q}$, where $2\leq q \leq5$. The simulation experiment parameters are: $\hbar$ = 3 (3 different programs running in the network: Twitter, Friendfeed, OS), $X$ = 10 (each program has 10 diversified implementations), $Q$ = 1 (every diversified implementation is vulnerable), $|\InitialCompromise|$ = 10 (10 programs/nodes are initially compromised), ${\cal F}_{\D,0}$ is our algorithm for employing initial diversity, $m_3$ = 5 (the attacker has 5 exploits against OS), $m_4$ = 10 (the attacker has 5 exploits against Twitter and 5 exploits against Friendfeed), and mission lifetime $T = 500$.

\begin{figure}[!htbp]
\vspace{-0.1cm}
\centering
\hspace{-0.5cm}
\subfloat[$\awd_{G,\hbar,X,Q}(\A,\D_3, 500)$\label{fig:rq4-1}]{\includegraphics[width=.54\linewidth]{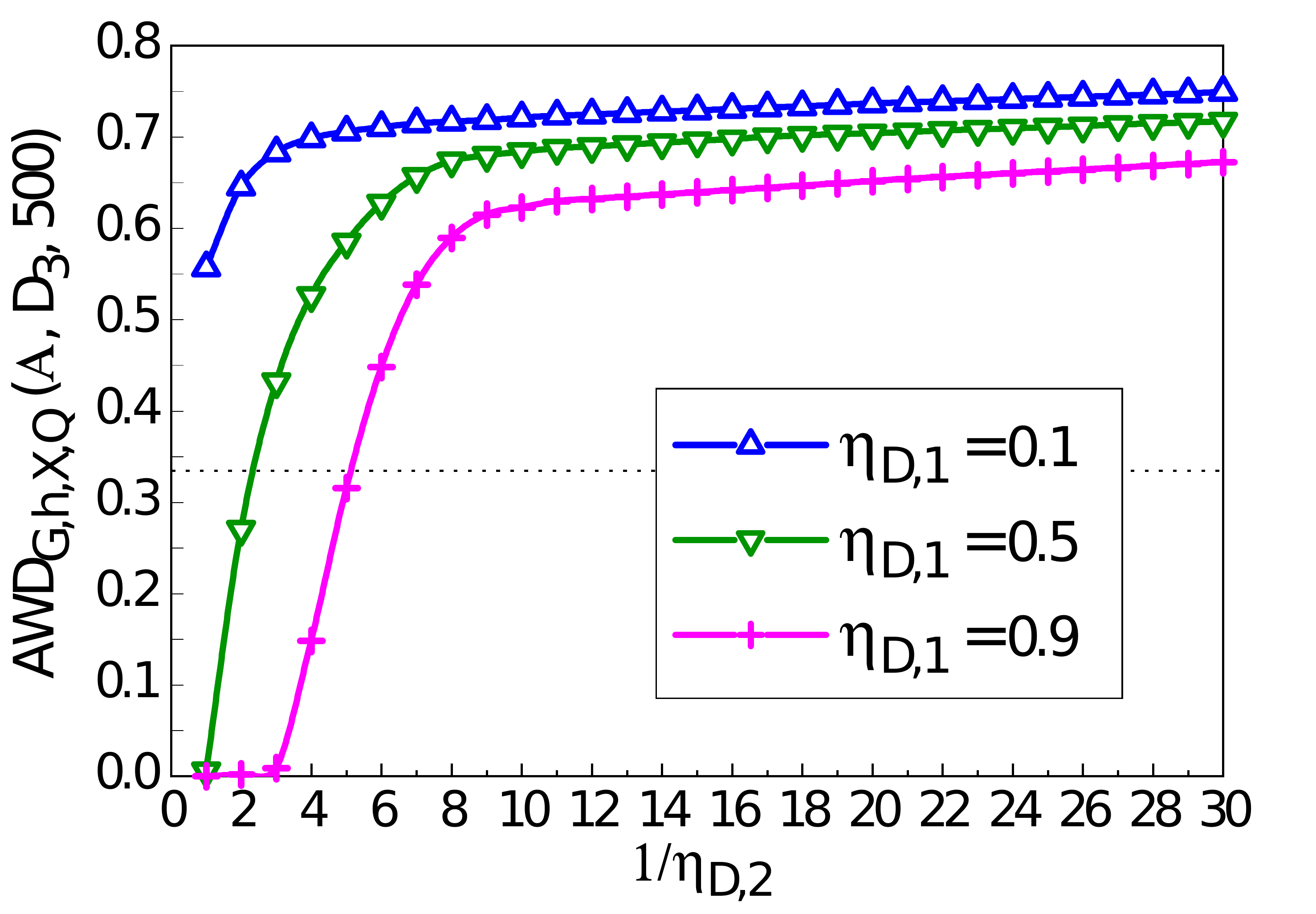}}
\hspace{-0.4cm}
\subfloat[$\aoc_{\D,3}^{\sf min}$\label{fig:rq4-2}]{\includegraphics[width=.54\linewidth]{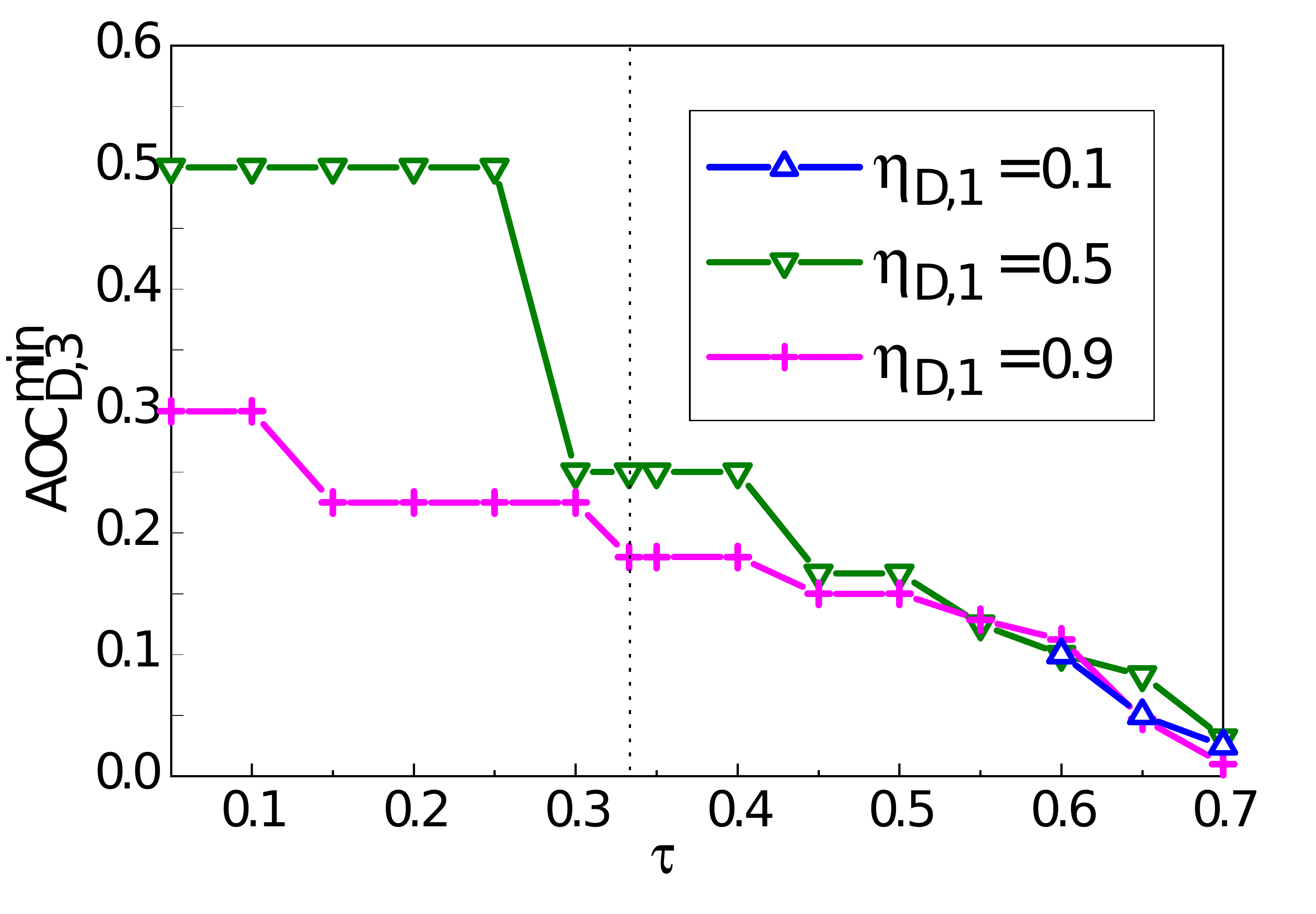}}
\hspace{-0.3cm}
\caption{\label{fig:rq4} Plots of {$\awd_{G,\hbar,X,Q}(\A,\D_3, 500)$} and $\aoc_{\D,3}^{\sf min}$.
}
\vspace{-0.1cm}
\end{figure}

Figure \subref*{fig:rq4-1} plots $\awd_{G,\hbar,X,Q}(\A,\D_3, 500)$ with respect to $\eta_{\D,1}$ and $1/\eta_{\D,2}$, where $\eta_{\D,1}$ is the the proportion of nodes where diversified programs are dynamically employed and $1/\eta_{\D,2}$ is the time interval between two consecutive diversity employments. We observe the following: (i) When $\eta_{\D,1}$ = 0.1, proactive diversity always leads to high attack worst damage, even if $1/\eta_{\D,2}$ is small (i.e., high-frequency in employment). This means that proactive diversity is useless when dynamic diversity is employed at a few nodes. (ii) When $\eta_{\D,1}$ = 0.5 or 0.9, proactive diversity can lead to low attack worst damage only when $1/\eta_{\D,2}$ is small. Consider defense goal $\tau$ =1/3 as an example, $1/\eta_{\D,2}$ must be no more than 2 when $\eta_{\D,1}$ = 0.5 and no more than 5 when $\eta_{\D,1}$ = 0.9. This means that proactive diversity is effective only when employed broadly and frequently. (iii) When $\eta_{\D,1}$ = 0.1 and $1/\eta_{\D,2} \geq 2$, or when $\eta_{\D,1}$ = 0.5 and $1/\eta_{\D,2} \geq 6$, or when $\eta_{\D,1}$ = 0.9 and $1/\eta_{\D,2} \geq 14$, proactive diversity leads to higher attack worst damage than static diversity. This means that proactive diversity can do more harm than good 
by making making more nodes exploitable over time.

\begin{insight}
Proactive diversity improves security only when employed at most nodes at high frequency.
\end{insight}

From Figure \subref*{fig:rq4-1}, we can derive the minimum average operational cost $\aoc_{\D,3}^{\sf min}$ with respect to the defender's goal $\tau\in[0,0.7]$, which is shown in Figure \subref*{fig:rq4-2}. Consider $\tau$ = 1/3 as an example, the minimum average operational cost is $\eta_{\D,1}\times \eta_{\D,2} = 0.5\times 1/2 = 0.25$ for $\eta_{\D,1}$ = 0.5, and $\eta_{\D,1}\times \eta_{\D,2} = 0.9\time 1/5 = 0.18$ for $\eta_{\D,1}$ = 0.9. Note that in the case $\eta_{\D,1}$ = 0.1, $\aoc_{\D,3}^{\sf min}$ is undefined when $\tau\in[0,0.55]$ because the strategy can never prevent the attacker from breaking the defender's goal. We further observe that $\eta_{\D,1}$ = 0.9 leads to a lower $\aoc_{\D,3}^{\sf min}$ than $\eta_{\D,1}$ = 0.5 when $0\leq\tau\leq 0.5$, meaning that a higher proportion of dynamic diversity re-employment leads to a lower average operation cost.

\begin{insight}
When proactive diversity is effective, {\color{black}a higher proportion of dynamic re-employment} leads to a lower operational cost than what is incurred by a higher re-employment frequency.
\end{insight}

\begin{figure}[!htbp]
\vspace{-0.3cm}
\centering
\hspace{-0.5cm}
\subfloat[$\awd_{G,\hbar,X,Q}(\A,\D_4, 500)$ \label{fig:rq4-3}]{\includegraphics[width=.54\linewidth]{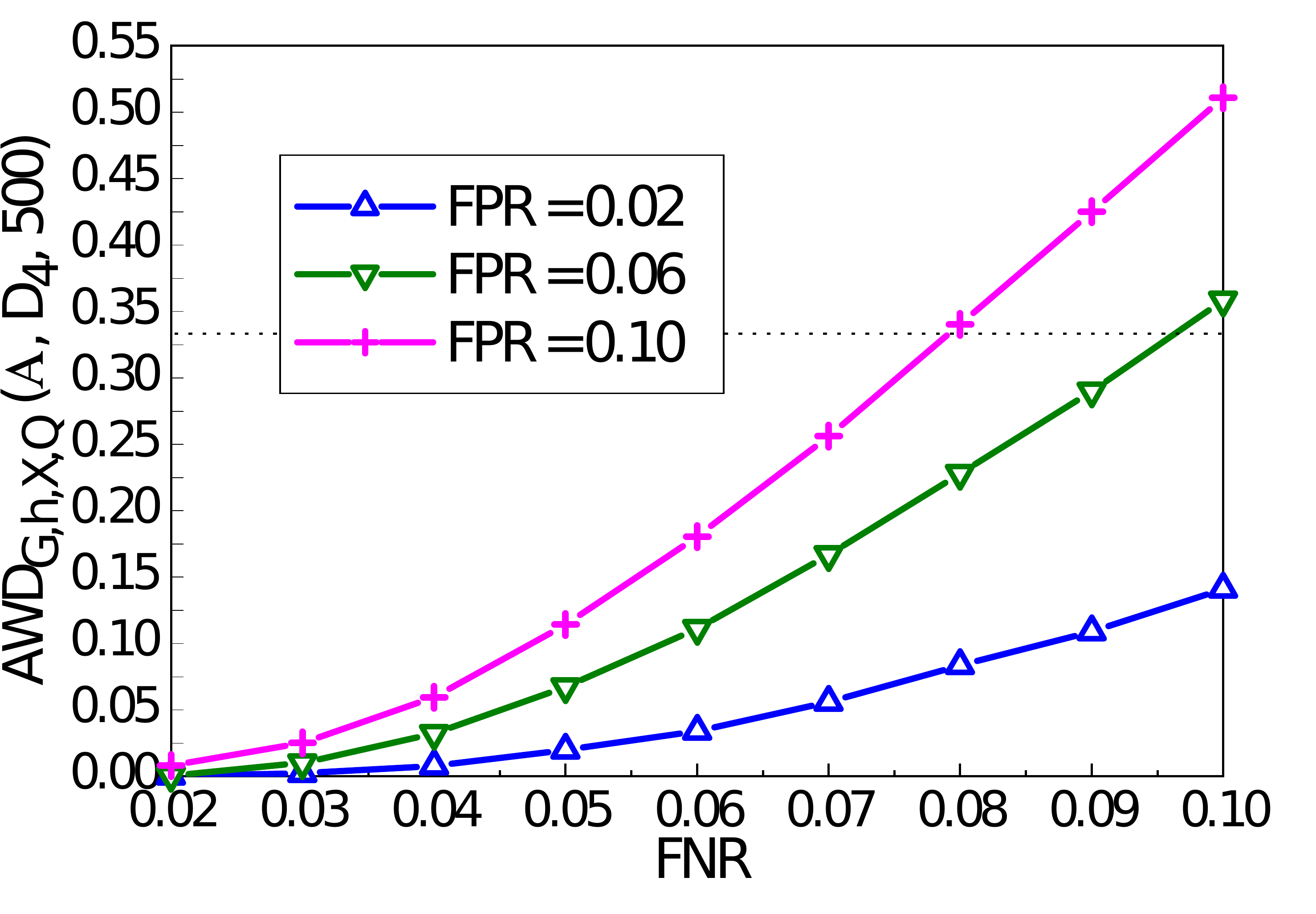}}
\hspace{-0.4cm}
\subfloat[$\aoc_{\D,4}$ \label{fig:rq4-4}]{\includegraphics[width=.54\linewidth]{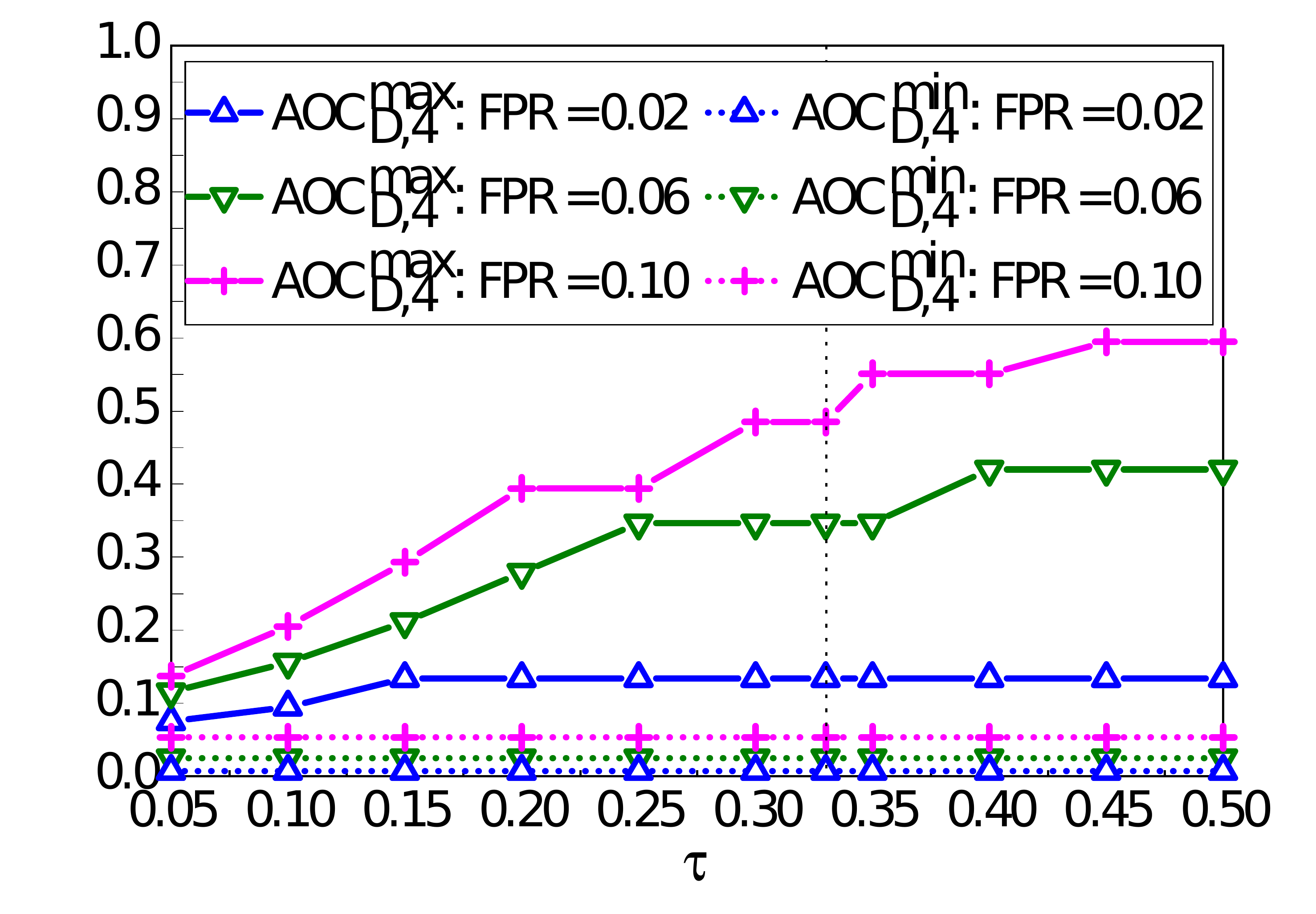}}
\hspace{-0.3cm}
\caption{\label{fig:rq4_2} Plots of $\awd_{G,\hbar,X,Q}(\A,\D_4, 500)$ and $\aoc_{\D,4}$.
}
\vspace{-0.1cm}
\end{figure}

Figure \subref*{fig:rq4-3} plots $\awd_{G,\hbar,X,Q}(\A,\D_4, 500)$ with respect to $\fpr$ (false-positive rate) and $\fnr$ (false-negative rate), where $\fpr,\fnr\in[0.02, 0.1]$. We observe that a lower $\fnr$ and $\fpr$ (i.e., higher attack-detection capability) leads to a lower attack worst damage, meaning the effectiveness of reactive-adaptive diversity largely depends on the attack-detection capability. Consider defender's mission goal of $\tau =1/3$ as an example, we observe that reactive-adaptive diversity can assure the mission when $\fpr = 0.02$ and $0.02\leq\fnr\leq0.1$, when $\fpr = 0.06$ and $0.02\leq\fnr\leq0.09$, and when $\fpr = 0.1$ and $0.02\leq\fnr\leq0.07$. From Figure \subref*{fig:rq4-3}, we can derive the $\aoc_{\D,4}^{\sf min}$ and $\aoc_{\D,4}^{\sf max}$ for a given $\fpr$  with respect to $\tau\in[0,0.5]$, which is plotted in Figure \subref*{fig:rq4-4}. We make the following observations. (i) For a fixed $\fpr$, $\aoc_{\D,4}^{\sf min}$ can be very low and remains stable as $\tau$ increases, because $\aoc_{\D,4}^{\sf min}$ is always achieved when $\fnr$ = 0.02. The operational cost is low because a high attack-detection accuracy can detect {\color{black}compromised computers} before they attack the others. (ii) For a fixed $\fpr$, a higher $\tau$ often leads to a higher $\aoc_{\D,4}^{\sf max}$, because a higher $\tau$ (i.e., higher compromise-tolerance) can be achieved at a lower operational cost from a diversity-based defense standpoint. (iii) The defender's operational cost falls into a wide range as $\tau$ increases, meaning that the defender's operational cost largely depends on the attack-detection effectiveness.

\begin{figure}[!htbp]
\vspace{-0.1cm}
\centering
\hspace{-0.5cm}
\subfloat[$\awd_{G,\hbar,X,Q}(\A,\D_5, 500)$ \label{fig:rq4-5}]{\includegraphics[width=.54\linewidth]{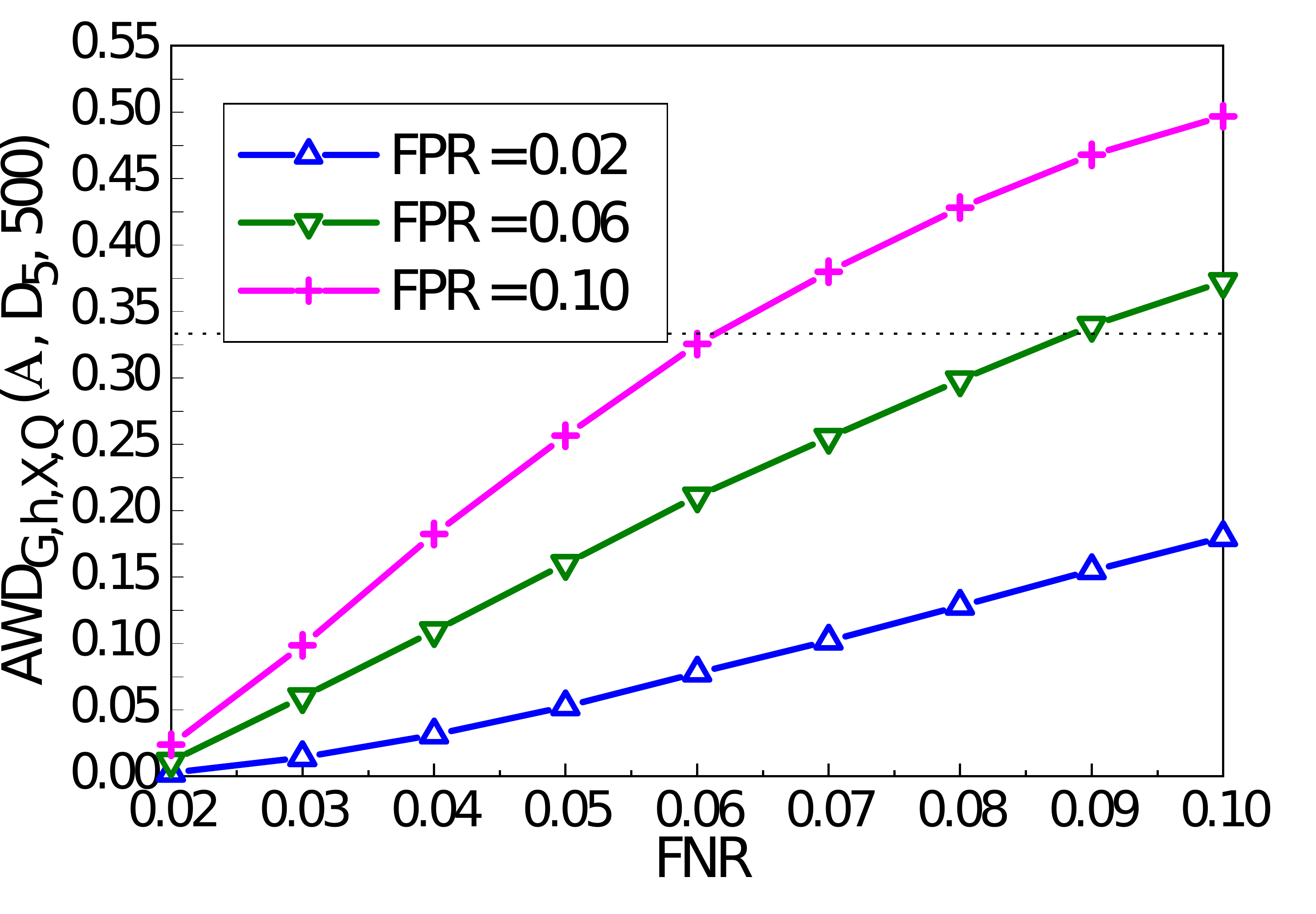}}
\hspace{-0.4cm}
\subfloat[$\aoc_{\D,5}$ \label{fig:rq4-6}]{\includegraphics[width=.54\linewidth]{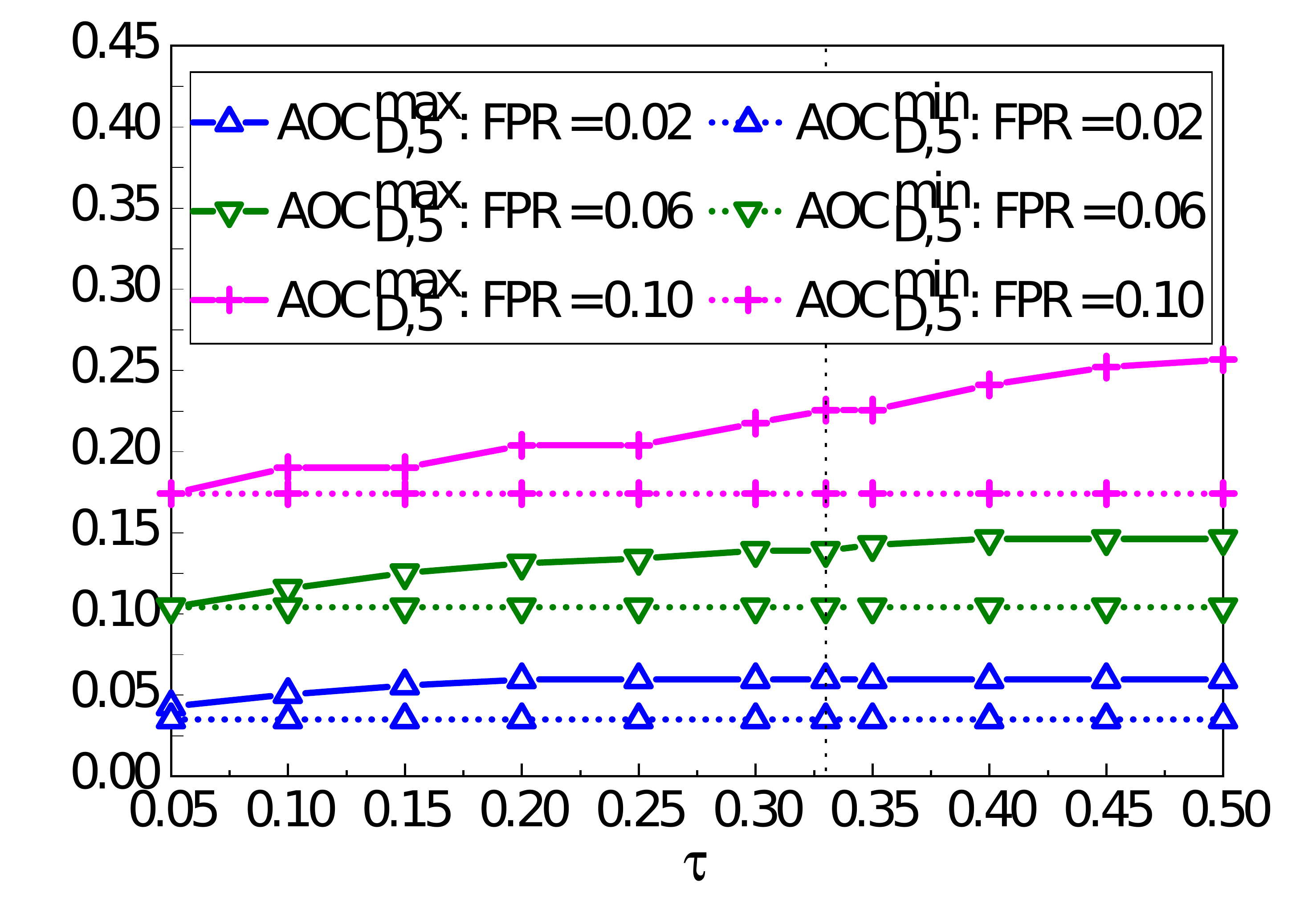}}
\hspace{-0.3cm}
\caption{\label{fig:rq4_3}Plots of $\awd_{G,\hbar,X,Q}(\A,\D_5, 500)$ and $\aoc_{\D,5}$.
}
\vspace{-0.1cm}
\end{figure}

Figure \subref*{fig:rq4-5} plots $\awd_{G,\hbar,X,Q}(\A,\D_5, 500)$ with respect to $\fpr$ and $\fnr$, where $\fpr,\fnr\in [0.02, 0.1]$ and re-deployment frequency $\eta_{\D,2} = 0.2$. From Figure \subref*{fig:rq4-5}, we derive $\aoc_{\D,5}^{\sf max}$ and $\aoc_{\D,5}^{\sf min}$, which is plotted in Figure \subref*{fig:rq4-6}. We observe that the defender's average operational cost is small. Consider $\tau$ = 1/3 as an example, we observe $\aoc_{\D,5}^{\sf min}= 0.0349$ when  $\fpr = 0.02$ and $\aoc_{\D,5}^{\sf max}=0.2254$ when $\fpr$ = 0.1, meaning $\aoc_{\D,5}\in$[0.0349, 0.2254] when $\fpr\in[0.02, 0.1]$ and $\fnr\in[0.02, 0.1]$.

By comparing Figures \subref*{fig:rq4-2}, \subref*{fig:rq4-4}, and \subref*{fig:rq4-6}, we observe that for $\tau\in[0.05,0.5]$, we have $\aoc_{\D,3}\in[0.15, 0.5]$ when $\eta_1 \in [0.5,0.9]$; we have $\aoc_{\D,4}\in[0.0071,0.5948]$ and $\aoc_{\D,5}\in[0.0349,0.2566]$ when $\fpr\in[0.02, 0.1]$ and $\fnr\in[0.02, 0.1]$, meaning that reactive-adaptive diversity incurs an average operational cost falling into a wider range than proactive diversity and an even wider range than hybrid diversity. {We also observe that proactive diversity incurs a higher operational cost than
reactive-adaptive diversity when the defender's tolerable compromise threshold $\tau$ is small, and the opposite is true when $\tau$ is large.}

\subsubsection{RQ5: Is it true that the more diversified implementations the better? }
In order to answer RQ5, we investigate how the {\em attacker slow-down} metric $\asd_{\D,q}$, the {\em attack extra cost} metric $\aec_{\D,q}$, and the {\em vulnerability tolerance} metric $\VT_{\D,q}$ depend on the number of diversified implementations $X$, where $2\leq q \leq5$. The experimental parameters are: $\hbar = 3$ (3 different programs running in the network: Twitter, Friendfeed, and OS), $Q = 1$ (every diversified implementation is vulnerable), $|\InitialCompromise| = 10$ (10 programs or nodes are initially compromised), $m_3$ = 1 (the attacker has 1 exploit against OS), $m_4$ = 2 (the attacker has 1 exploit against Twitter and 1 exploit against Friendfeed), ${\cal F}_{\D,0}$ is our algorithm for employing initial diversity, $\eta_{\D,1} = 0.5$ (diversified implementations are dynamically re-employed at 50\% of all nodes), $\eta_{\D,2} = 0.2$ (diversified implementations are re-employed every 5 time steps), $\fpr = 0.1$ (10\% false-positive rate in detecting attacks), $\fnr = 0.1$ (10\% false-negative rate in attack detection), and mission lifetime $T = 500$.

\begin{figure}[!htbp]
\vspace{-0.1cm}
\centering
\hspace{-0.5cm}
\subfloat[$\asd_{\D,q}$ \label{fig:rq5-1}]{\includegraphics[width=.54\linewidth]{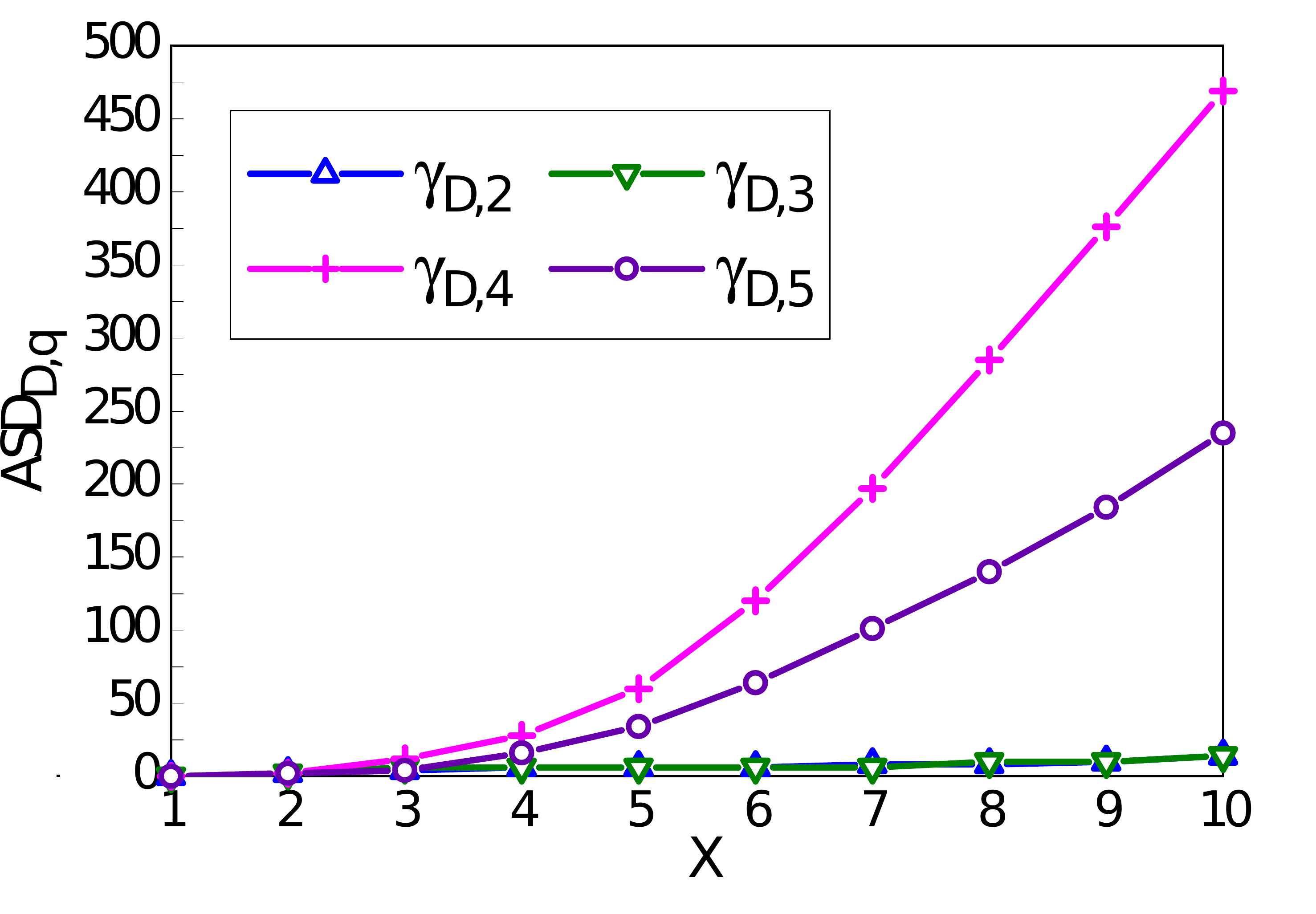}}
\hspace{-0.4cm}
\subfloat[$\aec_{\D,q}$ \label{fig:rq5-2}]{\includegraphics[width=.54\linewidth]{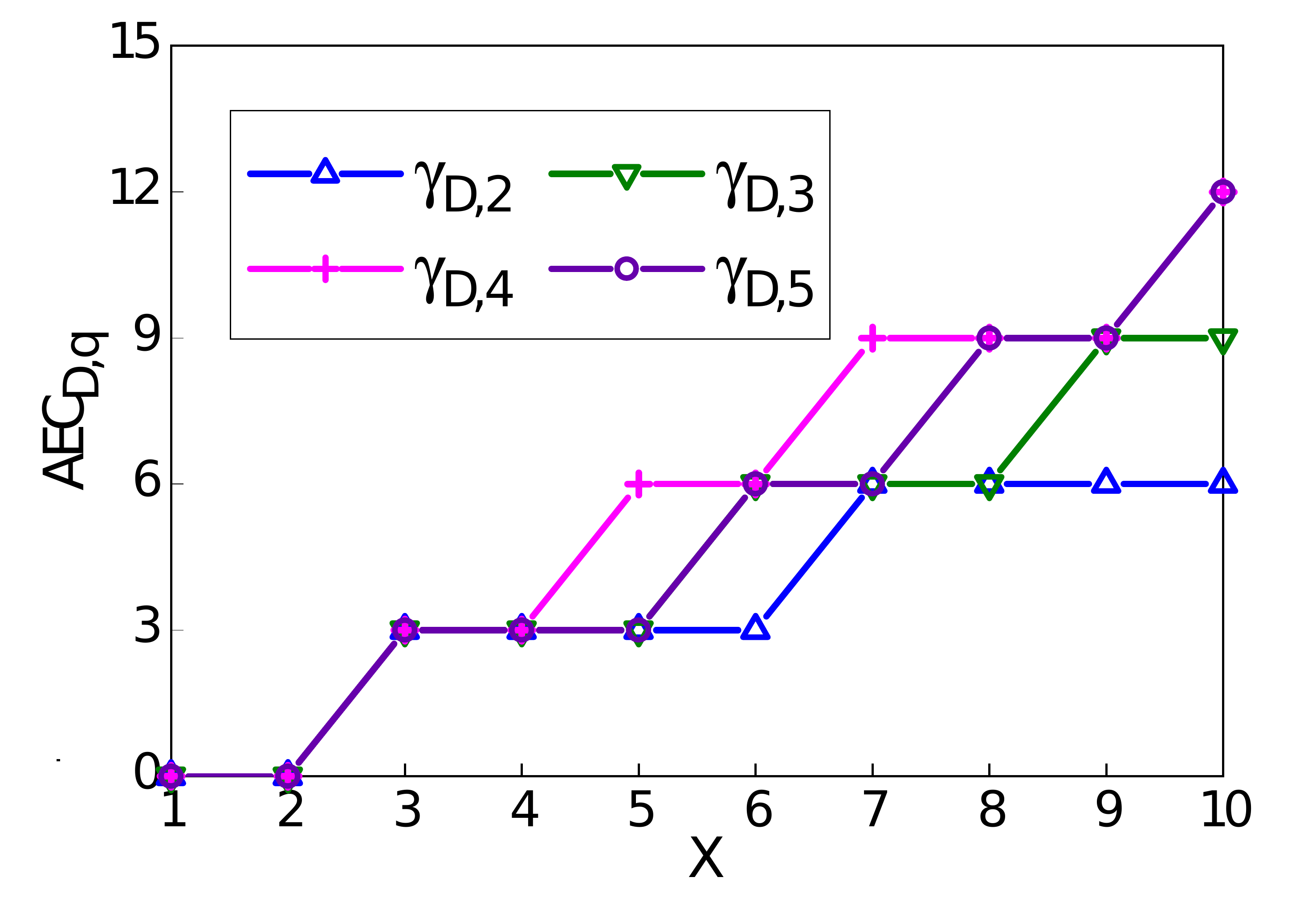}}
\hspace{-0.3cm}
\caption{Plots of $\asd_{\D,q}$ and $\aec_{\D,q}$, where $2\leq q \leq5$.}
\vspace{-0.1cm}
\end{figure}

Figure \subref*{fig:rq5-1} plots $\asd_{\D,q}$ with respect to $X$ and $\tau = 0.01$, where $2\leq q \leq5$. We take $\tau = 0.01$ as an example because the attacker possessing one exploit for each program may only compromise about 1\% computers during the mission lifetime $T=500$ with $X$ =10. We observe that $\asd_{\D,4}$ increases rapidly with $X$, followed by $\asd_{\D,5}$, while $\asd_{\D,3}$ and $\asd_{\D,2}$ increases slowly with $X$ (e.g., $\asd_{\D,3}$ = $\asd_{\D,2}$ = 2 when $X$ =2 and $\asd_{\D,3}$ = $\asd_{\D,2}$ = 14 when $X$ = 10). When $X$ = 10, reactive-adaptive diversity slows down the attacker most. 

Figure \subref*{fig:rq5-2} plots $\aec_{\D,q}$ with respect to $X$ and $\tau$ = 1/3, where $2\leq q \leq5$. We observe that $\aec_{\D,q}$ shows an upward trend as $X$ increases for $q$ = 2, 3, 4, 5, meaning that the more diversified implementations, the higher the attack extra cost for disrupting the defender's mission goal ({i.e., no more than a $\tau$ fraction of computers are compromised at any time $t\in [0,T]$}). In addition, we observe $\aec_{\D,4}\geq \aec_{\D,5}\geq\aec_{\D,3}\geq\aec_{\D,2}$ for any $X\in[1,10]$, meaning that reactive-adaptive diversity benefits most from more diversified implementations, followed by hybrid diversity, proactive diversity and static diversity.

In order to characterize the impact of the number of diversified implementations on the vulnerability tolerance $\VT_{\D,q}$ where $2\leq q \leq5$, we observe $\VT_{\D,q} = 0$ when $X = 1, 2$ and $\tau$ =1/3. This is reasonable because the attacker having one exploit for each program can easily break the defender's goal when the total number of diversified implementations is no more than 2. {In contrast, given $\tau$ =1/3, we have $\VT_{\D,q}$ = 1 when $X\geq3$, where $2\leq q \leq5$, because Figure \subref*{fig:rq5-2} shows that an attacker having one exploit for each program will never compromise 1/3 computers when $X\geq3$; this is true even if all of the diversified implementations are vulnerable. This means that when the attacker has a limited capability, a substantially higher vulnerability rate (than $\tau$) can effectively be tolerated until the number of diversified implementations reaches a certain amount.}

\begin{insight}
The more diversified implementations with a similar quality, the higher the attacker slow-down, the attack extra cost, and the vulnerability tolerance. 
\end{insight}

\section{Related Work}
\label{sec:related-work}

\noindent{\bf Prior studies in software diversity}. Prior studies mainly aim at obtaining diversified, ideally {\em independent}, implementations of a program specification \cite{baudry2015multiple}.
There is a body of literature on software diversity (e.g., \cite{chen1978n,DBLP:journals/tdsc/HomescuJCBLF17,DBLP:conf/ccs/Franz15,DBLP:conf/ndss/CraneHBLF15,DBLP:journals/ieeesp/LarsenBF14,bhatkar2003address,forrest1997building, xu2003transparent,kc2003countering,barrantes2003randomized}). 
However, the effectiveness of these building-block techniques is not well understood. For example, one study shows that independent software implementation does not lead to independent vulnerabilities because programmers tend to make the same mistakes \cite{knight1986experimental}. Other studies show positive results, such as: the same vulnerabilities in different software may demand different exploits \cite{han2009effectiveness}; few vulnerabilities simultaneously appear in different OSes \cite{henriques2019diverse}.
In contrast, the effectiveness of run-time diversity is better understood: 
(i) address space layout with 32-bit randomization can be compromised by brute-forcing \cite{shacham2004effectiveness}; (ii) address space layout with higher entropy can be compromised by side-channel attacks \cite{davi2015isomeron,evans2015missing,strackx2009breaking,DBLP:conf/ndss/RuddSBDHCLLDFSO17}; (iii) address space layout randomization is vulnerable because of modern cache architectures \cite{gras2017aslr}; (iv) fine-grained address space layout randomization is subject to just-in-time code reuse attacks  \cite{snow2013just}; and (v) instruction set randomization is subject to brute-forcing attacks
\cite{sovarel2005s,weiss2006known}.

The preceding studies consider standalone software diversification techniques. In contrast, we investigate the effectiveness of software diversification techniques from a network standpoint. Since we consider diversifying network-wide software stacks, multiple diversification techniques can be used together. This is reminiscent of the notion of $N$-variant systems \cite{cox2006n}, which however do not quantify the network-wide effectiveness dynamic diversity.

\smallskip

\noindent{\bf Prior studies in network diversity}. Network diversity has been investigated in some contexts
\cite{Geer2003,ODonnellCCS2004,Stamp:2004:RM:971617.971650,zhang2001heterogeneous,XuComplexNetworkSub2018}. 
There are proposals on measuring static network diversity via (i) the entropy of the distribution of software vulnerability in a network \cite{neti2012software} and (ii) the
diversity index of the shared vulnerabilities between different software implementations in a network \cite{temizkan2017software}.
There are attempts at optimizing static diversity by using some flavor of {\em graph coloring} algorithm (i.e., treating a diversified implementation as a color and a network as a graph) and minimizing defective edges (i.e., adjacent nodes have the same color or run the same software implementation) \cite{ODonnellCCS2004,huang2014toward,yang2008improving,XuComplexNetworkSub2018,10.1002/spe.2180,borbor2016diversifying}.

The closely related prior studies are \cite{wang2014modeling,zhang2016network},
which measure network diversity via {\em resource richness} and {\em attack effort}. Our study is different as follows: (i) they investigate how to quantify diversity, whereas we study how to employ {\em dynamic} diversity; (ii) they do not consider attack-defense interactions, whereas we explicitly model attack-defense interactions; (iii) we quantify the network-wide effectiveness of dynamic diversity, which is not studied by them. 

\smallskip

\noindent{\bf Prior studies related to Moving-Target Defense (MTD)}.
Proactive diversity is one form of MTD, which includes other kinds of proactive defense techniques (e.g., proactively changing IP addresses or port numbers) \cite{cho2020toward}. Quantifying the security effectiveness of MTD in the broader context is beyond the scope of the present paper (see, e.g., \cite{XuHotSOS14-MTD}). While our finding that reactive-adaptive is more effective than proactive diversity (i.e., MTD when applied to diversity) may sound counter-intuitive at a first glance, it can be understood as follows:
MTD (i.e., proactive diversity in this case) can be employed when the defender does not know the situation information of the network (e.g., which and how many nodes are compromised); in contrast, adaptive diversity can leverage the situational information to adaptively employ diversity, leading to potentially higher effectiveness.
In addition, our simulation study focuses on dynamic diversity against malware-like attacks after the attacker establishes footholds at some compromised computers in a network. This means that the simulation study does not consider earlier stages of cyber attacks, such as reconnaissance. Our findings do not contradict the usefulness of MTD in defending against such earlier-stage attack activities (e.g., MTD can effectively disrupt attacker's reconnaissance processes \cite{huang2011introducing,luo2014effectiveness,carroll2014analysis,jafarian2015effective}).

\smallskip

\noindent{\bf Prior studies related to whole-network security analysis}. We analyze the security effectiveness of dynamic diversity from a whole-network perspective. There are studies in this perspective, but tackling different problems and using different approaches. The {\em attack graph} approach studies how an attacker may exploit multiple vulnerabilities to achieve a certain goal and how to harden a network (see, e.g.,  \cite{phillips1998graph,sheyner2002automated,ritchey2000using,albanese2012time,ammann2002scalable, cheng2014metrics,homer2013aggregating,zhang2018network,wang2018network}). This approach is {\em combinatorial} in nature and does not consider the {\em time} dimension \cite{XuMTD2020,xu2019cybersecurity,XuCybersecurityDynamicsHotSoS2014}. Another approach is the {\em cybersecurity dynamics} framework \cite{XuMTD2020, xu2019cybersecurity,XuCybersecurityDynamicsHotSoS2014}, which explicitly models attack-defense interactions over time and includes a rich family of models and results (e.g., \cite{XuTDSC2011,XuTAAS2012,XuTDSC2012,XuInternetMath2012,XuGameSec13,XuHotSOS14-MTD,Cam2014,XuTAAS2014,XuHotSoS2015,XuIEEETNSE2018,DBLP:journals/tnse/YangYT18,XuIEEEACMToN2019}).
These studies aim at analytical results while making simplifying assumptions, such as the {\em independence} assumption between attacks. In order to eliminate such assumptions, initial efforts have been made in both theoretical studies  \cite{XuInternetMath2012,XuQuantitativeSecurityHotSoS2014,XuInternetMath2015Dependence}
and empirical studies  \cite{XuHotSoS2018Firewall,XuHotSoS2018Diversity}. Our framework does not make the independence assumption, while characterizing the {\em transient} behaviors (i.e., the dynamics before converging to an equilibrium); whereas, analytical models so far can only offer  asymptotic results (i.e., $t\to \infty$ or when the dynamics converges to the equilibrium). Last but not the least, our framework goes much beyond \cite{XuHotSoS2018Diversity}
by characterizing (e.g.) dynamic attack-defense interactions and decision-making.

\section{Limitations}
\label{sec:limitations}

The present study has several limitations that need to be addressed in future research. 
First, the framework has two limitations. (i) 
We assume $G$ is fixed. This implicitly assumes the network defense tools are not compromised because {\color{black}a successful attack against} a defense tool can effectively change $G$. While this is reasonable for missions with a short lifetime $T$, it is interesting to accommodate dynamic $G_t$ for missions of long lifetime $T$ and the case that the network defense tools can be compromised, as outlined in \cite{XuMTD2020,xu2019cybersecurity,XuCybersecurityDynamicsHotSoS2014}. (ii) We assume that the attacker selects one tool to use at each phase of an attack strategy. This can be extended to using multiple tools in a sequential manner. 

Second, the simulation study has some limitations.
(i) We only consider simple decision-making algorithms, which are sufficient for demonstrating the usefulness of the framework but need to accommodate more sophisticated decision-making algorithms. 
(ii) We use some ``synthetic'' and simplifying scenarios owing to the lack of real data, meaning that the findings may not be generalized to other scenarios. Specifically, the assumption that each attack phase takes place at one time step may limit the validity of Insight \ref{insight:RQ1}; the assumption that each exploit incurs the same cost to the attacker may limit the validity of Insight \ref{insight:RQ2};
the assumption that each implementation of a program is equally vulnerable may limit the validity of Insight \ref{insight:RQ3}; the independence assumption that different implementations do not have common vulnerabilities holds in some settings \cite{han2009effectiveness,garcia2011diversity} but may not hold in general. 
In order to see the potential impact of these assumptions, we conduct additional experiments where different implementations can contain common vulnerabilities. While omitting the experimental details owing to space limit, they do show that the independence assumption does cause overestimates of the effectiveness of employing network diversity. This resonates the results of earlier theoretical studies \cite{XuInternetMath2012,XuQuantitativeSecurityHotSoS2014,XuInternetMath2015Dependence}.

\section{Conclusion}
\label{sec:conclusion}

We proposed a framework for quantifying the cybersecurity effectiveness of enforcing (dynamic) network diversity, including a suite of security metrics for measuring attacker's cost (incurred by obtaining exploits) and defender's operational cost (incurred by re-employing network diversity). We conducted simulation experiments to measure these metrics with respect to a number of dynamic diversity strategies and drew insights from the experimental results. 

There are many open problems. In addition to those mentioned in Section \ref{sec:limitations}, we highlight two: How can we obtain analytic results without making strong assumptions?
How can we quantify attack power and defense power without making strong assumptions?

\smallskip

\noindent{\bf Acknowledgement}. We thank the reviewers for their constructive comments that guided us in revising the paper. We thank Eric Ficke for proofreading the paper. We thank Dr. Alexander Kott for illuminating discussions. This work was supported in part by ARO Grant \#W911NF-17-1-0566,
NSF Grants \#2115134 and \#2122631 (\#1814825), and by a Grant from the State of Colorado.


\begin{IEEEbiography}[{\includegraphics[width=1in,height=1.25in,clip,keepaspectratio]{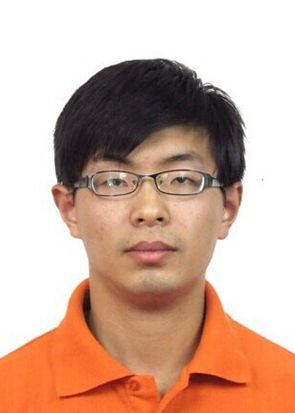}}]
{Huashan Chen} received the M.S. degree from the Institute of Information Engineering, Chinese Academy of Sciences, in 2016. He is currently pursuing the Ph.D. degree with the Department of Computer Science, University of Texas at San Antonio. His primary research interests are in cybersecurity, especially moving target defense and security metrics.
\end{IEEEbiography}
\vspace{-12mm}
\begin{IEEEbiography}[{\includegraphics[width=1in,height=1.25in,clip,keepaspectratio]{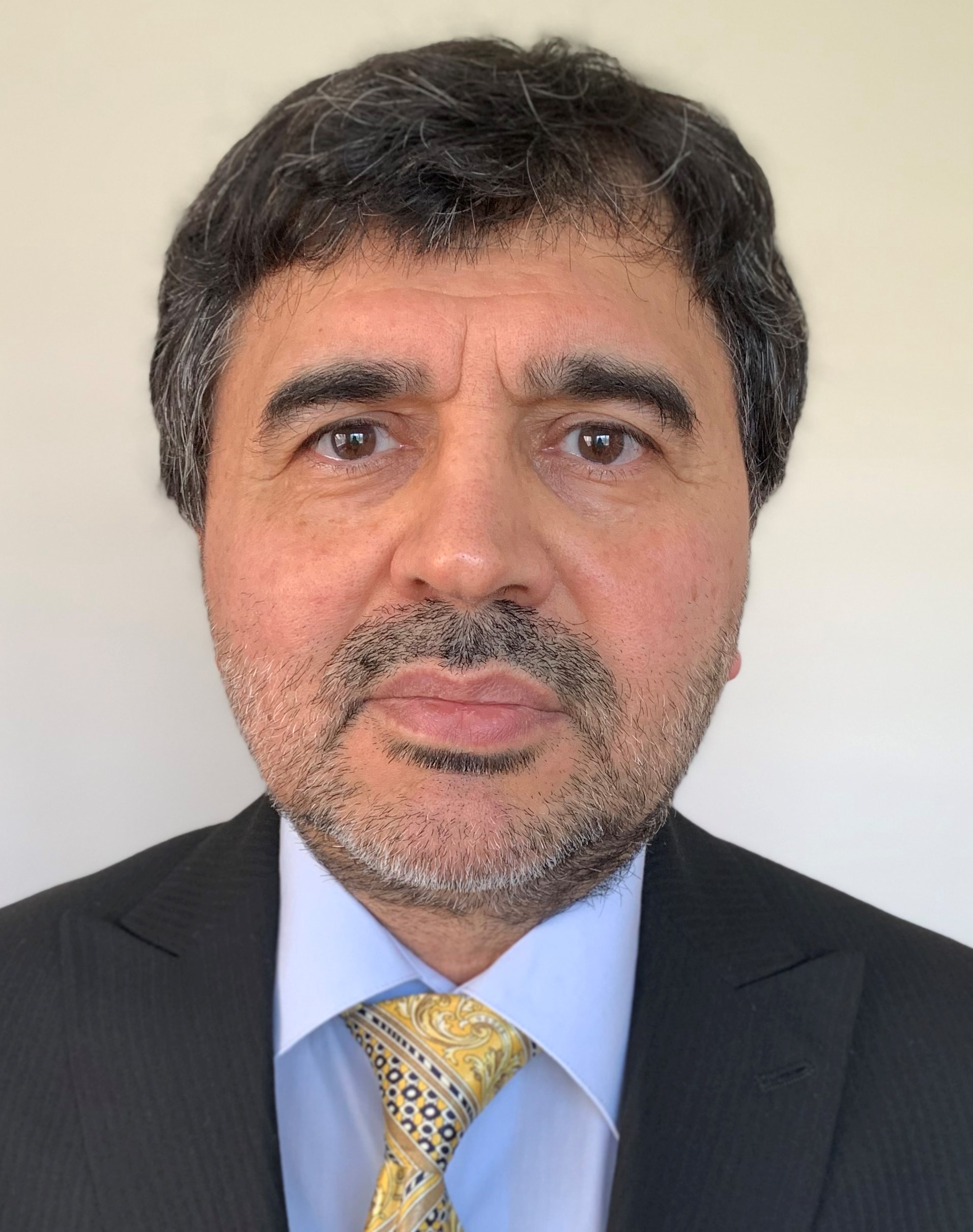}}]
{Hasan Cam} is a Principal Machine Learning Scientist at Best Buy in USA. His current research interests include machine learning, cybersecurity, data analytics, autonomous agents, swarm intelligence, and networks. He currently works on the projects involved with applying machine learning and graph techniques to anomaly and fraud detection. Prior to joining Best Buy, he was a Computer Scientist at the Army Research Laboratory, where he served as the government lead for the Risk area in Cyber Collaborative Research Alliance. Previously, he also worked as a faculty member in the academia and a senior research scientist in the industry. He has published more than 35 journal and 70 conference papers, while having two patents and several patent applications. He has served as an editorial member of two journals, a guest editor of two special issues of journals, an organizer of symposiums and workshops, and a Technical Program Committee Member in numerous conferences. He received the Ph.D. degree in electrical and computer engineering from Purdue University, and the M.S. degree in computer science from Polytechnic University, New York. He is a Senior Member of IEEE.
\end{IEEEbiography}
\vspace{-12mm}
\begin{IEEEbiography}
[{\includegraphics[width=1in,height=1.25in,clip,keepaspectratio]{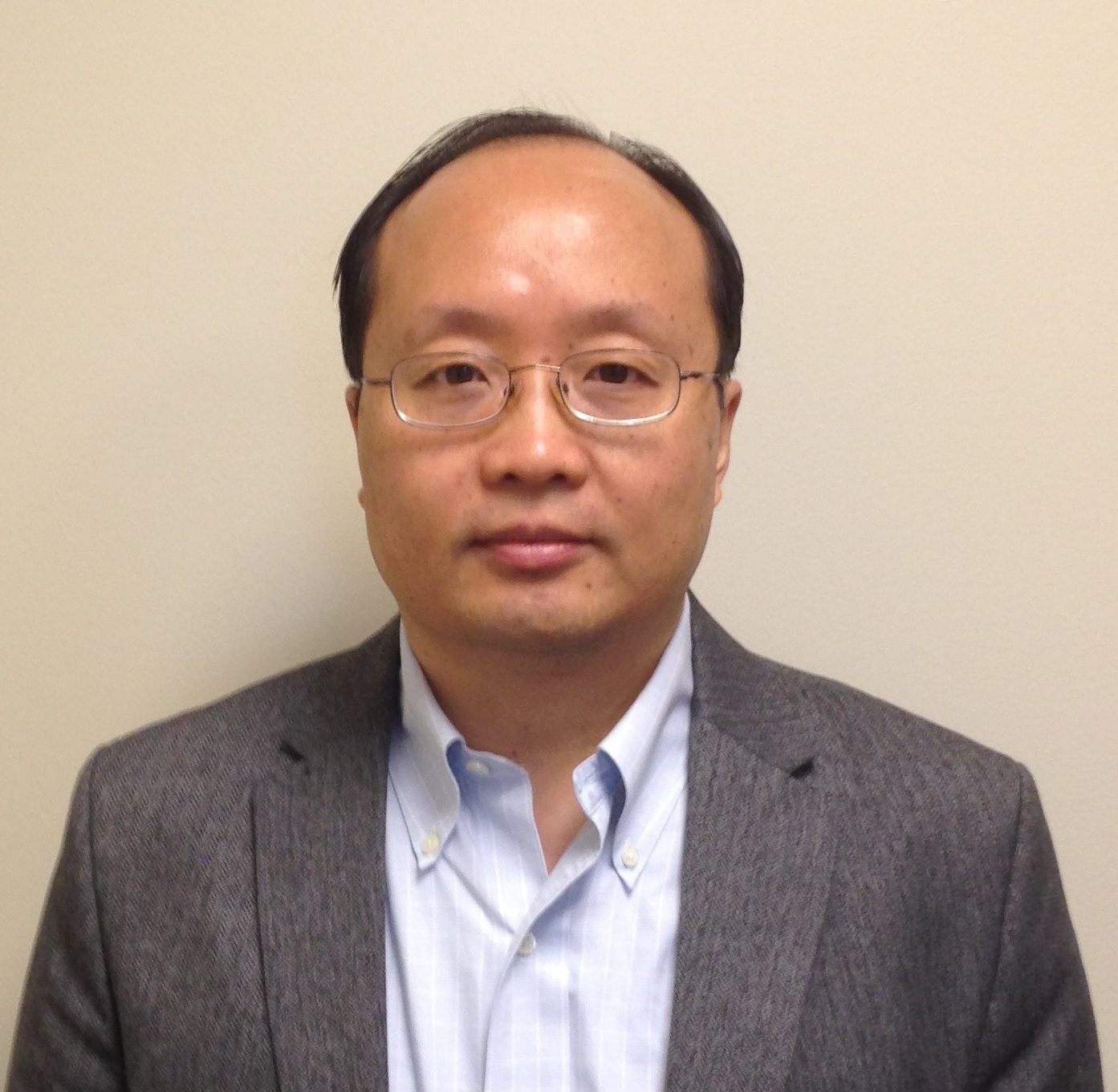}}]
{Shouhuai Xu}
(M\textquoteright14\textendash SM\textquoteright20)
is the Gallogly Chair Professor in the Department of Computer Science, University of Colorado Colorado Springs (UCCS). Prior to joining UCCS, he was with University of Texas at San Antonio. He pioneered the Cybersecurity Dynamics approach as foundation of the emerging science of cybersecurity, with three pillars: first-principle cybersecurity modeling and analysis (the $x$-axis); cybersecurity data analytics (the $y$-axis); and cybersecurity metrics (the $z$-axis, to which the present study belongs).  He co-initiated the International Conference on Science of Cyber Security and is serving as its Steering Committee Chair. He is/was an Associate Editor of IEEE Transactions on Dependable and Secure Computing (IEEE TDSC), IEEE Transactions on Information Forensics and Security (IEEE T-IFS), and IEEE Transactions on Network Science and Engineering (IEEE TNSE). He received a PhD degree in Computer Science from Fudan University.
\end{IEEEbiography}

\end{document}